\documentclass[12pt]{article}

\usepackage[fleqn]{amsmath}
\usepackage{amsfonts}
\usepackage{amssymb}
\usepackage{mathrsfs}

\usepackage[dvipsnames]{xcolor}

\usepackage{epsfig}

\usepackage{caption}
\usepackage{float}
\usepackage[title]{appendix}

\usepackage{hyperref}
\hypersetup{
    colorlinks,
    citecolor=blue,
    filecolor=blue,
    linkcolor=blue,
    urlcolor=blue
}

\newcommand{\NN}{\mathbb{N}}
\newcommand{\RR}{\mathbb{R}}
\newcommand{\SP}{\mathbb{S}}
\newcommand{\ZZ}{\mathbb{Z}}

\newcommand{\ONE}{\boldsymbol{1}}

\newcommand{\beg}{\begin{equation}}
\newcommand{\en}{\end{equation}}

\newcommand{\eps}{\varepsilon}

\DeclareFontFamily{U}{mathx}{}
\DeclareFontShape{U}{mathx}{m}{n}{<-> mathx10}{}
\DeclareSymbolFont{mathx}{U}{mathx}{m}{n}

\DeclareMathAccent{\widecheck}{0}{mathx}{"71}

\begin{document}

\title{Bounds on $T_c$ in the Eliashberg theory of\\ Superconductivity. II: Dispersive phonons\vspace{-0.5truecm}}
\author{M. K.-H. Kiessling,$^1\!$ B. L. Altshuler,$^2\!$  and E. A. Yuzbashyan$^3$\\ \small
          $^1$ Department of Mathematics,\\ \small
Rutgers, The State University of New Jersey,\\ \small
          110 Frelinghuysen Road, Piscataway, NJ 08854\\ \small
          $^2$ Department of Physics, \\ \small
Columbia University,\\ \small
             538 West 120th Street, New York, NY 10027\\ \small
             $^3$ Center of Materials Theory, \\ \small Department of Physics and Astronomy, \\ \small
Rutgers, The State University of New Jersey,\\ \small
          136 Frelinghuysen Road, Piscataway, NJ 08854}

\date{Revised version of July 12, 2025}
\maketitle

\thispagestyle{empty}

\vfill\vfill
\hrule
\medskip
\noindent
\copyright(2025) 
\small{The authors. Reproduction of this preprint, in its entirety, is permitted for non-commercial purposes only.}

\newpage
\abstract{\noindent
 The standard Eliashberg theory of superconductivity is studied, in which 
the effective electron-electron interactions are modelled as mediated by generally dispersive phonons,
with Eliashberg spectral function $\alpha^2\!F(\omega)\geq 0$ that is $\propto\omega^2$ for small $\omega>0$
and vanishes for large $\omega$.
 The Eliashberg function also defines the electron-phonon coupling strength 
$\lambda:= 2 \displaystyle\int_{\RR_+}\!\! \frac{\alpha^2\!F(\omega)}{\omega}d\omega$.
 Setting ${ \displaystyle\frac{2\alpha^2\!F(\omega)}{\omega}}d\omega =: \lambda P(d\omega)$, 
formally defining a probability measure $P(d\omega)$ with compact support, and assuming as usual that the phase transition between 
normal and superconductivity coincides with the linear stability boundary $\mathscr{S}_{\!c}$ of the normal 
region in the $(\lambda,P,T)$ parameter space against perturbations toward the superconducting region, 
it is shown that this \textsl{critical hypersurface} $\mathscr{S}_{\!c}$ is a graph of a function $\Lambda(P,T)$.
 This proves that the normal and the superconducting regions are simply connected.
 Moreover, it is shown that $\mathscr{S}_{\!c}$ is determined by a variational principle: 
if $(\lambda,P,T)\in\mathscr{S}_{\!c}$, then $\lambda = 1/\mathfrak{k}(P,T)$, where 
$\mathfrak{k}(P,T)>0$ is the largest eigenvalue of a compact self-adjoint operator $\mathfrak{K}(P,T)$
on $\ell^2$ sequences that is constructed explicitly in the paper, for all admissible $P$. 
 Furthermore, given any such $P$, sufficient conditions on $T$ are stated under which the map 
$T\mapsto \lambda = \Lambda(P,T)$ is invertible.
 For sufficiently large $\lambda$ this yields the following:
(i) the existence of a critical temperature $T_c$ as function of $\lambda$ and $P$;
(ii) an ordered sequence of lower bounds on $T_c(\lambda,P)$ that converges to $T_c(\lambda,P)$.
 Also obtained is an upper bound on $T_c(\lambda,P)$.
 Although not optimal, it agrees with the asymptotic 
 form $T_c(\lambda,P) \sim C \sqrt{\langle \omega^2\rangle} \sqrt{\lambda}$ valid for $\lambda\sim\infty$, given $P$, 
though with a constant $C$ that is a factor $\approx 2.034$ larger than the sharp constant;
here, $\langle\omega^2\rangle := \int_{\RR_+} \omega^2 P(d\omega)$.}

\newpage

\section{Introduction}\label{sec:INTRO}\vspace{-.2truecm}

 This is our second paper in a series of papers in which we rigorously inquire into the phase transition 
between {the normal and superconducting states} as accounted for by the Eliashberg theory \cite{migdal,Eliashberg,AllenMitrovic,Carbotte,AllenDynes,Marsiglio}.
 Each paper uses results of the previous ones, while addressing a problem that is of interest in its own right. 

 In \cite{KAYgamma}, the first paper of our series, to which we refer the reader also for a ``master introduction'' 
to the whole project, we studied a version of Eliashberg theory known as the $\gamma$ model,
introduced  recently in \cite{MoonChubukov} (see also \cite{ChubukovETal}), which aims at describing superconductivity in
systems close to a quantum phase transition, where the effective electron-electron interactions are mediated 
by collective bosonic excitations (fluctuations of the order parameter field) instead of phonons.
 The $\gamma$ model contains three parameters: $T$, $\gamma$, and $g$; yet $g$ enters only
in the combination $T/g$, so that effectively the model contains only two parameters.
 We were able to characterize the critical temperature $T_c$ in terms of a variational principle valid for all $\gamma>0$, viz.
\begin{equation}
\label{eq:gammaVP}
T_c(g,\gamma) = \tfrac{g}{2\pi} \big[\mathfrak{g}(\gamma)\big]^\frac1\gamma,
\end{equation}
where $\mathfrak{g}(\gamma)>0$ is the largest eigenvalue of a self-adjoint, compact operator $\mathfrak{G}(\gamma)$
constructed in \cite{KAYgamma}.
 Based on this variational principle we obtained a sequence of lower bounds on $T_c$ that converges upward to $T_c$.
 The $N$-th lower bound is defined by the largest zero of a polynomial of degree $N$.
 The first four of these lower bounds we computed explicitly in terms of elementary functions; the bounds for $N>4$ {generically} 
cannot be expressed in this way, as per Galois theory.
 Plotting our bounds jointly into a diagram suggests that for $\gamma \geq 2$ the fourth lower approximant to $T_c$ 
should be accurate enough for most practical purposes.
 So far, however, the $\gamma$ model has been applied in the theory of superconducting materials only with {certain values of $\gamma<2$ 
including $\gamma\in\{\frac13,\frac12,\frac{7}{10},1\}$; see  \cite{ChubukovETal}}.
 For these $\gamma$ values an accurate computation of $T_c(g,\gamma)$ requires a numerical approximation
of $\mathfrak{g}(\gamma)$ with rank $N>4$ approximations to $\mathfrak{G}(\gamma)$.

 At $\gamma=2$ the $\gamma$ model captures the asymptotic behavior at infinite electron-phonon coupling strength $\lambda$ of the
standard Eliashberg theory in which the effective electron-electron interactions are mediated by generally dispersive phonons,
with a spectral function\footnote{We remark that $\alpha^2\!F$ is a compound symbol.
      This standard notation has historical roots.  For further comments, see \cite{KAYgamma}.}
$\alpha^2\!F(\omega)\geq 0,$ {known as the Eliashberg function,} that is $\propto\omega^2$ for small $\omega>0$
and vanishes for large $\omega$.
 The spectral function also defines the electron-phonon coupling strength 
$\lambda:= 2 \int_{\RR_+} \frac{\alpha^2\!F(\omega)}{\omega}d\omega$.
 This standard version of Eliashberg theory has successfully explained  {the critical temperature $T_c$ and other properties of most conventional superconductors \cite{AllenMitrovic,Carbotte,AllenDynes,TroyanETal}.}

 In the present paper we study this standard version of Eliashberg theory.
 Setting $2\alpha^2\!F(\omega)/\omega=: \lambda P^\prime(\omega)$, with $P^\prime(\omega)$ the density with respect to
Lebesgue measure of a formal probability measure $P(d\omega)$ that behaves $\propto \omega$ for small $\omega>0$, and 
whose support is contained in a bounded interval $[0,\overline{\Omega}(P)]$ with $0< \overline{\Omega}(P)$, and assuming as 
usual that the phase transition between normal and superconductivity coincides with the linear stability boundary 
$\mathscr{S}_{\!c}$ of the normal region in the positive $(\lambda,P,T)$ cone against {perturbations
toward the superconducting region}, 
we will show that this \textsl{critical hypersurface} $\mathscr{S}_{\!c}$ is a graph of over the positive $(P,T)$ cone.
 Hence the normal and the superconducting regions in the positive $(\lambda,P,T)$ cone are each simply connected.
 Moreover, we will show that the critical surface is determined by a variational principle: 

\textsl{If $(\lambda,P,T)\in\mathscr{S}_{\!c}$, then}
\begin{equation}
\label{eq:lambdaVP}
\lambda = 1\big/\mathfrak{k}(P,T),
\end{equation}

\textsl{where $\mathfrak{k}(P,T)\!>\!0$ is the largest eigenvalue of a self-adjoint}

\textsl{compact operator $\mathfrak{K}(P,T)$ on $\ell^2$ sequences that is explicitly}

\textsl{constructed in {section~\ref{sec:THMoneANDtwoPROOFS} of this} paper, for all admissible $P$.} 

 Approximating $\mathfrak{K}(P,T)$ with a nested sequence of finite-rank operators that converges to $\mathfrak{K}(P,T)$,
an increasing sequence of rigorous lower bounds on $\mathfrak{k}(P,T)$ is obtained
that converges to $\mathfrak{k}(P,T)$; these lower bounds on $\mathfrak{k}(P,T)$ translate into upper bounds
on $\lambda\in \mathscr{S}_c$, given $P$.
 The first four of these can be, and are computed explicitly in closed form, involving a handful of expected values w.r.t. $P$.
 Also an explicit rigorous upper bound on $\mathfrak{k}(P,T)$ is obtained, which translates into a rigorous lower bound on
$\lambda\in \mathscr{S}_c$ for each $(P,T)$.
 Furthermore, conditioned on $P$ being given, we state sufficient conditions on $T$ for the map $T\mapsto \lambda$ 
and its lower $N$-frequency approximations to be invertible.
 For sufficiently large $\lambda$  this yields the following:

 (i) the existence of a critical temperature $T_c(\lambda,P)$;

 (ii) a sequence of lower bounds on $T_c(\lambda,P)$ converging to $T_c(\lambda,P)$.
\newpage

\noindent
 Also obtained is an upper bound on $T_c(\lambda,P)$, which is not optimal yet agrees with the asymptotic 
behavior $T_c(\lambda,P) \sim C \sqrt{\langle \omega^2\rangle} \sqrt{\lambda}$ for large enough $\lambda$, given $P$, 
though with $C\approx 2.034 C_\infty$, where $C_\infty = 0.1827262477...$ is the sharp constant.
 Here, $\langle\omega^2\rangle := \int_{\RR_+} \omega^2 P(d\omega)$.

  Several results about dispersive phonons we will be able to prove by reduction to the proofs of analogous results for
the $\gamma$ model in \cite{KAYgamma}, or by suitable adaptations of these proofs.
 Yet some results obtained in the present paper required totally new arguments.

 The results of the present paper are for a rather general class of models, and therefore less quantitative {than} those
in \cite{KAYgamma}. 
 Quantitative results analogous to those obtained in \cite{KAYgamma}  require a specification of
$2\alpha^2\! F(\omega)$, something that has to be left to studies that are aimed at concrete applications.
 One such specific model is the non-dispersive limit of the standard Eliashberg theory, in which
\begin{equation}\label{eq:EinsteinAlphaF}
 2\alpha^2\! F(\omega)\big|^{}_{\mbox{\rm{\tiny{E}}}} := \lambda\Omega\delta (\omega-\Omega);
\end{equation}
here, $\Omega$ is the Einstein frequency of the phonons. 
 This model will be evaluated in \cite{KAYeinstein}.

 In the remaining sections we first precisely specify the basic technical quantities of the Eliashberg theory.
 We employ its recent reformulation in terms of a classical Bloch spin chain model \cite{YuzAltPRB}.
 We then list our main results, followed by several sections with their proofs.

\section{The Bloch spin-chain model}\label{sec:spinCHAIN}
\vspace{-5pt}

 As in \cite{KAYgamma}, we  work with the \textsl{condensation energy} of Eliashberg theory, 
the difference between the grand (Landau) potentials of the superconducting and normal states.
 Using units where Boltzmann's constant $k_{\mbox{\tiny{B}}}=1$ and the reduced Planck constant $\hbar =1$,
its spin chain representation reads (cf. \cite{KAYgamma}, eq.(33))
\begin{align}\label{eq:H}
 {H}({\bf S}|{\bf N}) : = &2\pi 
\sum\limits_{n} \omega_n {\bf N}_0\cdot\big({\bf N}_n -{\bf S}_n \big)
  \\
\notag & + \pi^2 {T}
          \sum\!\sum\limits_{\hskip-0.4truecm n\neq m}
          \lambda_{n,m}^{} \left({\bf N}_n\cdot {\bf N}_m -{\bf S}_n\cdot{\bf S}_m \right).
\end{align}
 Here, ${\bf S}_n \in {\SP}^1\subset {\RR}^2$ with $n\in{\ZZ}$ 
denotes the $n$-th spin in the Bloch spin chain ${\bf S}\in ({\SP}^1)^{\ZZ}$.
 The spin chain ${\bf N}$ is associated with the {\it{normal state}} of the Migdal--Eliashberg theory, 
having $n$-th spin given by ${\bf N}_n := -{\bf N}_0\in  {\SP}^1\subset {\RR}^2$ for $n < 0$ and 
${\bf N}_n:={\bf N}_0$ for $n\geq 0$.
 Any other \textsl{admissible} spin chain satisfies the asymptotic conditions that, sufficiently fast,
${\bf S}_n \to {\bf N}_n$ when $n\to \infty$ and when $n\to - \infty$, where ``sufficiently fast'' is 
explained below.
 Moreover, admissible spin chains satisfy the symmetry relationship {that for all $n\in{\ZZ}$, 
 ${\bf N}_0\cdot {\bf S}_{-n} = - {\bf N}_0\cdot{\bf S}_{n-1}$ and 
 ${\bf K}_0\cdot {\bf S}_{-n} =  {\bf K}_0\cdot{\bf S}_{n-1}$, 
where ${\bf K}_0\in {\SP}^1\subset {\RR}^2$ is an arbitrary vector perpendicular to ${\bf N}_0$.
 This} reduces the problem to effective spin chains ${\bf S}\in  ({\SP}^1)^{{\NN}_0}$, with ${\NN}_0:={\NN}\cup\{0\}$.
 The central dots between two such spins indicate Euclidean dot product, with ${\bf S}_n$ and ${\bf N}_n$ 
understood as vectors in ${\RR}^2$ of unit length.

 Furthermore, $\lambda_{n,m}^{}\equiv\lambda(n-m)$ in (\ref{eq:H}) is a (dimensionless) positive spin-pair interaction kernel,
which for the standard Eliashberg theory with dispersive phonons reads (cf. \cite{KAYgamma}, eq.(7))
\begin{equation}\label{eq:lambdaGENERAL}
\lambda(n-m)
:= 2 \int_0^\infty \frac{\alpha^2\! F(\omega)\omega}{\omega^2+(\omega_n-\omega_m)^2} d\omega,
\end{equation}
where the $\omega_n := 2n\pi T$ are the Matsubara frequencies, and where 
$\omega\mapsto \alpha^2\!F(\omega)\in {\cal L}^1_+(\RR_+, d\omega)$ is the {Eliashberg} electron-phonon spectral function.
 It is defined as the thermodynamic limit of a finite-volume-sample sum of Dirac $\delta$ measures with positive coefficients, 
concentrated on a discrete set of frequencies $\Omega_k\in [\underline{\Omega},\overline{\Omega}]$ 
with $0<\underline{\Omega}\leq\overline{\Omega}$, which converges in weak$^*$ topology to a function 
$\alpha^2\!F(\omega)\propto\omega^2$ for small $\omega$, yet still vanishing for $\omega>\overline{\Omega}$. 
 Note that $\lambda(n-m) = \lambda(m-n)$, and that for each $k\in\NN_0$ the expression $\lambda(k)$ is a $T$-dependend
bounded linear functional of $\alpha^2\!F$; the dependence on $\alpha^2\!F$ and on $T$ of all $\lambda(\,\cdot\,)$ is not exhibited explicitly.

 With this convention, the dimensionless electron-phonon coupling constant of the theory  $\lambda\equiv \lambda(0)$.
 Explicitly,
\begin{equation}\label{eq:lambda}
\lambda = 2 \int_0^\infty \frac{\alpha^2\! F(\omega)}{\omega} d\omega.
\end{equation}
 Note that our $\lambda$ is the \textit{standard} (renormalized) dimensionless electron-phonon coupling constant 
of the Eliashberg theory; cf. \cite{AllenDynes}.

 We have collected all the information needed to define \textsl{admissibility} of a spin chain ${\bf S}$ to mean,
that after employing their stipulated symmetry to convert all summations over negative Matsubara frequencies 
in the sum over $\ZZ$ and the double sum over $\ZZ^2$ in (\ref{eq:H}) into summations over positive ones,
the so rewritten (\ref{eq:H}) converges absolutely.

 Not every admissible spin chain qualifies as thermal equilibrium state. 
 For a spin chain to actually represent a thermal equilibrium state it needs to minimize the condensation 
energy functional (\ref{eq:H}).
\smallskip

\noindent
{\bf Conjecture~1}: \textsl{
There is a critical temperature $T_c >0$, depending on $\alpha^2\!F(\omega)$,
such that for temperatures  $T\geq T_c$, the spin chain of the normal state $\mathbf{N}$ is the unique
minimizer of $H({\mathbf S}|{\mathbf N})$, whereas at temperatures $T<T_c$ a spin chain ${\bf S}\neq {\bf N}$ 
for a {\it{superconducting phase}} minimizes $H({\mathbf S}|{\mathbf N})$ uniquely 
up to an irrelevant gauge transformation (fixing of an overall phase).
 Moreover, the phase transition at $T_c$ from normal to superconductivity is continuous.} 
 \smallskip 

 In this paper we take some steps toward the rigorous vindication of this conjecture.
 As in \cite{KAYgamma}, we will \textsl{assume} the existence of 
a continuous phase transition between normal and superconductivity,
so that its location in the phase diagram coincides with the linear-stability boundary of the 
normal state against small perturbations {toward the superconducting region, from now on
referred to as ``superconducting perturbations.''}
 Thus we will rigorously study the Eliashberg gap equations linearized about the normal state.
 A confirmation of the continuous normal-to-superconductivity phase transition requires
a study of the nonlinear Eliashberg gap equations, which we postpone to a later publication.

 We next summarize our main results.
\vspace{-20pt}

\section{Main results}\label{sec:mainRESULTS}
\vspace{-10pt}

 Since it has become customary to emphasize the dependence of the critical temperature on the electron-phonon coupling constant 
$\lambda \equiv \lambda(0)$, we rewrite the effective electron-electron interaction mediated by generally dispersive phonons as 
\vspace{-0.2truecm}
\begin{equation}\label{eq:VphononGENERALalt}
\lambda(n-m)
=: \lambda \int_0^\infty \frac{ \omega^2}{\omega^2+(\omega_n-\omega_m)^2} P(d\omega),
\end{equation}

\vspace{-0.2truecm}
\noindent
 where $P(d\omega)$ is a measure that integrates to unity, 
with a density $P^\prime(\omega)$ w.r.t. Lebesgue measure that behaves $\propto\omega$ for small $\omega$ 
and vanishes for $\omega>\overline{\Omega}(P)$.
 We denote the set of these measures by ${\mathcal{P}}$.

\smallskip
\noindent
{\bf Theorem~1}: 
\textit{The positive $(\lambda,P,T)$ cone of the standard Eliashberg model consists of two simply connected regions.
 In one region the normal state is unstable against small superconducting perturbations, in the other region it is
linearly stable.
 The boundary between the two regions, called the critical hypersurface $\mathscr{S}_{\!c}$, is a graph over the set
$\{(P,T)\}$, i.e. $\mathscr{S}_{\!c}=\{(\lambda,P,T): \lambda = \Lambda(P,T)\}$. 
 The function $\Lambda$ is continuous in both variables.
 The thermal equilibrium state at temperature $T$ of a 
crystal with normalized phonon spectral density $P^\prime$ and electron-phonon coupling constant $\lambda$ is
the superconducting phase when $\lambda>\Lambda(P,T)$ and the normal (metallic) phase when  $\lambda<\Lambda(P,T)$.}
\newpage

 Our Theorem~1 does not rule out that some lines 
$\mathscr{L}(\lambda,P):=\{(\lambda,P,T): \lambda \ \&\ P\ \mbox{fixed}\}$ could pierce $\mathscr{S}_{\!c}$ more 
than once, in which case the critical surface would not be a graph over the set $\{(\lambda,P)\}$
of the crystal model parameters --- at odds with Conjecture~1. 
 To rigorously confirm Conjecture~1's empirical thermodynamic narrative for the Eliashberg model, still \textsl{assuming} the existence
of a continuous normal-to-superconducting phase transition, one needs to show that $\Lambda(P,T)$ depends 
strictly monotonically on $T$, given $P$.
 We have the following result.

\smallskip
\noindent
{\bf Theorem~2}: 
\textit{The map $T\mapsto\Lambda(P,T)$ is strictly monotonic increasing on $[T_*,\infty)$, 
with $T_*(P) \leq {\overline\Omega(P)}/{2\sqrt{2}\pi}$.
 Hence, with $\lambda_*(P):=\Lambda(P,T_*(P))$, the portion of the critical surface $\mathscr{S}_{\!c}$ 
over the region $\{\lambda \geq \lambda_*(P)\}$ in the set of crystal parameters $(\lambda,P)$ is also a graph, 
yielding the critical temperature $T_c(\lambda,P)$, viz.}

\vspace{-.5truecm}
\begin{equation}
 \mathscr{S}_{\!c}\big|_{\lambda\geq\lambda_*}^{} = 
\big\{(\lambda,P,T): T = T_c(\lambda,P), \lambda\geq\lambda_*(P) \big\}.
\end{equation}
\smallskip

\noindent
{\bf Remark~1}.
\textit{In \cite{KAYgamma} we characterized $T_c(g,\gamma)$ explicitly in terms of the variational principle (\ref{eq:gammaVP}).
 In (\ref{eq:gammaVP}), $\mathfrak{g}(\gamma)>0$ is the largest eigenvalue of a self-adjoint, compact operator $\mathfrak{G}(\gamma)$
explicitly constructed in \cite{KAYgamma}.
 Based on this variational principle we were able to obtain lower bounds on $T_c(g,\gamma)$, four of these in closed form. 
 Our sequence of lower bounds on $T_c(g,\gamma)$ is expressed in terms of the largest eigenvalues of a sequence
of nested finite-rank approximations to the compact operator $\mathfrak{G}(\gamma)$ that converges monotonically
to $\mathfrak{G}(\gamma)$ when the rank $N$ is increased to $\infty$.
 With the help of Maple and, independently, Mathematica, we found that when the rank is increased beyond $N=200$, then
10 significant decimal places of the lower approximation to $T_c(g,\gamma=2)$ have stabilized, yielding
 $\frac1g T_c(g,2) = 0.1827262477...$.}
\smallskip

 For the standard Eliashberg model we have not been able to express $T_c(\lambda,P)$ as
(some power) of an eigenvalue of a suitable self-adjoint operator. 
 However, we are able to characterize $\Lambda(P,T)$ in terms of a variational principle that is mathematically similar
to how we characterized $T_c(g,\gamma)$ in the $\gamma$ model.

\smallskip
\noindent
{\bf Theorem~3}: 
\textit{The function $\Lambda(P,T)$ is determined by the following variational principle, 
\begin{equation}\label{eq:LambdaVP}
 \Lambda(P,T) = \frac{1}{\mathfrak{k}(P,T)}, 
\end{equation}
where ${\mathfrak{k}(P,T)}>0$ is the largest eigenvalue of {a
compact self-adjoint operator $\mathfrak{K}(P,T)$ on the Hilbert space of square-summable
sequences over the non-negative integers, with $\mathfrak{K}(P,T)$
explicitly constructed in section~\ref{sec:THMoneANDtwoPROOFS}, see (\ref{eq:QofXIrew})--(\ref{eq:HopsCompsTHREE}).}}
\smallskip

 Since compact operators on separable Hilbert spaces can be arbitrarily closely approximated by their finite-rank
approximations obtained by a sequence of restriction to finite-dimensional subspaces that converge monotonically 
to the Hilbert space, our variational principle~(\ref{eq:LambdaVP}) furnishes a sequence of upper approximations
to $\Lambda(P,T)$ that converges monotonically downward to $\Lambda(P,T)$. 
 The first four of these can be computed in closed form, using only elementary functions.

 More precisely, we have the following:
\smallskip

\noindent
{\bf Theorem~4}: \textsl{For all $N\in\NN$, $\Lambda(P,T) < 1/\mathfrak{k}^{(N)}(P,T)$, where 
$\mathfrak{k}^{(N)}(P,T)$ is the largest eigenvalue of $\mathfrak{K}^{(N)}(P,T)$, the 
restriction of $\mathfrak{K}(P,T)$ to the first $N$ components of $\ell^2({\NN}_0)$.
 The eigenvalues $\mathfrak{k}^{(N)}(P,T)$ can be explicitly computed for $N\in\{1,2,3,4\}$.
 They read:}
\begin{equation}\label{eq:kONE}
\mathfrak{k}^{(1)}(P,T)= \int_0^\infty 
\frac{\omega^2}{\omega^2+ (2\pi T)^2}P(d\omega),
\end{equation}
\textsl{which is the sole eigenvalue of $\mathfrak{K}^{(1)}(P,T)$;}
\begin{equation}\label{eq:kTWO}
\mathfrak{k}^{(2)}(P,T) =
\tfrac12\Big({\rm tr}\,\mathfrak{K}^{(2)} + \sqrt{\big({\rm tr}\,\mathfrak{K}^{(2)}\big)^2-4 \det\mathfrak{K}^{(2)}}\,\Big)(P,T),
\end{equation}
\textsl{where $\mathfrak{K}^{(2)}(P,T)$ is the upper leftmost $2\times2$ block of the matrix $\mathfrak{K}^{(4)}(P,T)$
displayed further below;}
\begin{align}\label{eq:kTHREE}
\hspace{-0.9truecm}
\mathfrak{k}^{(3)}(P,T)  = 
\tfrac13\!\left(\! \textstyle{ {\rm tr}\,\mathfrak{K}^{(3)} +
6\sqrt{\frac{p}{3}}\cos \left[\frac13\arccos\left(\frac{q}{2}\sqrt{\!\Big(\frac3p\Big)^{\!{}_3}}\,\right)\!\right] 
} \right)\!\!(P,T)
,\!\!\!
\end{align}
\textsl{with  (temporarily suspending displaying the dependence on $P,T$)}
\begin{align}\label{eq:3x3hP}
p = \tfrac13\!\big({\rm tr}\,\mathfrak{K}^{(3)}\big)^2- {\rm tr\, adj}\, \mathfrak{K}^{(3)},
\end{align}
\begin{align}\label{eq:3x3hQ}
q = 
 \tfrac{2}{27}\big({\rm tr}\,\mathfrak{K}^{(3)}\big)^3 
- \tfrac13 \big({\rm tr}\,\mathfrak{K}^{(3)}\big) \big({\rm tr\, adj}\,\mathfrak{K}^{(3)}\big)
+ \det\mathfrak{K}^{(3)},
\end{align}
\textsl{where $\mathfrak{K}^{(3)}(P,T)$ is the upper leftmost $3\times3$ block of the matrix $\mathfrak{K}^{(4)}(P,T)$
displayed further below;}
\begin{align}\label{eq:kFOUR}
\hspace{-0.8truecm}
\mathfrak{k}^{(4)}(P,T)  = 
\Big[\! \sqrt{\tfrac12 Z}
+\!\sqrt{\!\tfrac{3}{16} A^2 -\tfrac12 B - \tfrac12 Z +\tfrac{A^3 -4AB + 8C}{16\sqrt{2Z}}}\!-\tfrac14 A \Big]\!(P,T),
\end{align}
\textsl{where}
\begin{equation}
\label{eq:Zdef}
Z = \tfrac13 \Big[ \sqrt{Y} \cos\Big(\tfrac13 \arccos \tfrac{X}{2\sqrt{Y^3}}\Big) - B + \tfrac38 A^2\Big],
\end{equation}
\textsl{is a positive zero of the so-called \textsl{resolvent cubic} associated with the characteristic polynomial 
$\det \big(\eta\mathfrak{I} - \mathfrak{K}^{(4)}(P,T)\big)$, and}
\begin{align}
\label{eq:Xdef}
X =&  2B^3 - 9ABC+27C^2 + 27A^2D -72BD,\\
\label{eq:Ydef}
Y =& B^2-3AC+12D,
\end{align}
\textsl{where}
\begin{align}
\label{eq:Adef}
A &= - {\rm tr}\, \mathfrak{K}^{(4)},\\
\label{eq:Bdef}
B &= \tfrac12\Big( \big({\rm tr}\, \mathfrak{K}^{(4)}\big)^2 - {\rm tr}\, \big({\mathfrak{K}^{(4)}}\big)^2\Big),\\
\label{eq:Cdef}
C &= -\tfrac16 \Big( 
\big({\rm tr}\, \mathfrak{K}^{(4)}\big)^3 - 3\, {\rm tr}\, \big({\mathfrak{K}^{(4)}}\big)^2 \big({\rm tr}\, \mathfrak{K}^{(4)}\big)
+ 2\,{\rm tr}\, \big({\mathfrak{K}^{(4)}}\big)^3\Big),\\
\label{eq:Ddef}
D &= \det {\mathfrak{K}^{(4)}},
\end{align}
\textsl{with $\mathfrak{K}^{(4)}(P,T)=\int_0^\infty 
\mathfrak{H}^{(4)}\big(\varpi\big) P(d\omega)$, where we introduced the abbreviation $\varpi:={\omega}/{2\pi T}$, and where} 
\begin{align}\label{eq:Hfour}
& 
\mathfrak{H}^{(4)}= \\ \notag
&\hspace{-2.5truecm}  
{\begin{pmatrix}
{[\![}1{]\!]} 
&  \frac{1}{\sqrt{3}}\bigl({[\![}2{]\!]}+{[\![}1{]\!]}
\bigr)
&  \frac{1}{\sqrt{5}}\bigl({[\![}3{]\!]} + {[\![}2{]\!]}
\bigr)  
&  \frac{1}{\sqrt{7}}\bigl({[\![}4{]\!]}+{[\![}3{]\!]}
\bigr)   \\
 \frac{1}{\sqrt{3}}\bigl({[\![}2{]\!]}+{[\![}1{]\!]}
\bigr)
&  \frac{1}{3}\bigl({[\![}3{]\!]}-2{[\![}1{]\!]}
\bigr)
&  \frac{1}{\sqrt{15}}\bigl({[\![}4{]\!]}+{[\![}1{]\!]}
\bigr) 
&  \frac{1}{\sqrt{21}}\bigl({[\![}5{]\!]}+{[\![}2{]\!]}
 \bigr)  \\ 
 \frac{1}{\sqrt{5}}\bigl({[\![}3{]\!]}+{[\![}2{]\!]}
\bigr) 
&  \frac{1}{\sqrt{15}}\bigl({[\![}4{]\!]}+{[\![}1{]\!]}
\bigr) 
&  \frac{1}{5}\bigl({[\![}5{]\!]}-2({[\![}2{]\!]}+{[\![}1{]\!]})
\bigr) 
&  \frac{1}{\sqrt{35}}\bigl({[\![}6{]\!]}+{[\![}1{]\!]}
\bigr)  \\
 \frac{1}{\sqrt{7}}\bigl({[\![}4{]\!]}+{[\![}3{]\!]}
\bigr)  & 
 \frac{1}{\sqrt{21}}\bigl({[\![}5{]\!]}+{[\![}2{]\!]}
\bigr) & 
  \frac{1}{\sqrt{35}}\bigl({[\![}6{]\!]}+{[\![}1{]\!]}
\bigr)  & 
 \frac{1}{7}\bigl({[\![}7{]\!]}-2({[\![}3{]\!]}+{[\![}2{]\!]}+{[\![}1{]\!]})
\bigr) 
 \\
\end{pmatrix}}\!,
\end{align}
\textsl{with} ${[\![}n{]\!]}\big(\varpi\big):= \frac{\varpi^2}{n^2+ \varpi^2}\equiv \frac{\omega^2}{(2n \pi T)^2 +\omega^2}$ \textsl{for} $n\in\NN$.
\smallskip

\noindent
{\bf Remark~2}.
\textit{We have stated our Theorem~4 in a way that makes it obvious that the matrices $\mathfrak{K}^{(N)}(P,T)$ are
averages w.r.t. $P(d\omega)$ of matrices $\mathfrak{H}^{(N)}(\varpi)$ of non-dispersive models, with $\varpi={\omega}/{2\pi T}$.
 However, with the sole exception of the eigenvalue $\mathfrak{k}^{(1)}(P,T)$ of the first approximation, all eigenvalues
$\mathfrak{k}^{(N)}(P,T)$ for $N>1$ are not simply 
averages w.r.t. $P(d\omega)$ of the pertinent largest eigenvalues $\mathfrak{h}^{(N)}(\varpi)$ of $\mathfrak{H}^{(N)}(\varpi)$.
 The same is true for $\mathfrak{k}(P,T)$ in relation to $\mathfrak{h}(\varpi)$.
 Each $\mathfrak{k}^{(N)}(P,T)$ with $N\in\NN$ is a function of averages w.r.t. $P(d\omega)$ of 
the expressions $\frac{\omega^2}{\omega^2+ 4n^2\pi^2T^2}$ with $n\in\{1,...,2N-1\}$.}
\smallskip

 It is manifestly obvious that our lower bound $\mathfrak{k}^{(1)}(P,T)$ is strictly monotonic decreasing with $T\in\RR_+$,
given $P$. In section \ref{sec:TcTWO}
 of this paper strict decrease with $T\in\RR_+$ will be proved also for $\mathfrak{k}^{(2)}(P,T)$;
see Proposition~9.
  While we have not succeeded in showing that all the maps $T\mapsto\mathfrak{k}^{(N)}(P,T)$ and the map $T\mapsto\mathfrak{k}(P,T)$
are strictly monotonic decreasing for all $T\in\RR_+$, given $P$, we succeeded in proving decrease for sufficiently large $T$.
 More precisely, we have the following.
\smallskip

\noindent
{\bf Proposition~1}:\hspace{-2pt} 
\textsl{For $N\in\NN$ the eigenvalues $\mathfrak{k}^{(N)}(P,T)$ decrease strictly monotonically with $T\in[T_*(P),\infty)$.
 Moreover, $T_*(P)\leq {\overline\Omega(P)}/{2\sqrt{2}\pi}$.}
\smallskip

Our Theorem~2 is a consequence of Theorems~3 and~4, and of Proposition~1.
\smallskip

 As to the small-$T$ behavior of our upper bounds $\Lambda^{(N)}(P,T)$ to $\Lambda(P,T)$, we have proved that
in the limit $T\to 0$ the sequence of upper bounds $\Lambda^{(N)}(P,T)$ to $\Lambda(P,T)$ meets the $\lambda$ axis at explicitly 
computable $P$-independent locations $\lambda_N$ that converge slowly to $0$ like $1/\ln N$ as $N\to\infty$.
 More precisely, we have:
\smallskip

\noindent
{\bf Theorem~5}:\hspace{-3pt}
\textsl{The eigenvalues $\mathfrak{k}^{(N)}(P,T)$ converge when $T\to 0$ to $P$-independent numbers, viz.}
\begin{equation}\label{eq:lambdaN}
\lim_{T\to 0} \mathfrak{k}^{(N)}(P,T) = -1 + 2{\textstyle\sum\limits_{n=0}^{N-1} \frac{1}{2n+1}}
=: \mathfrak{k}^{(N)}_0.
\end{equation}
\textsl{Thus, as $T\to 0$ the $N$-th upper approximation $\Lambda^{(N)}(P,T)$ to $\Lambda(P,T)$ 
converges downward to $1/\mathfrak{k}^{(N)}_0=: \lambda_N$.
 Moreover, $\mathfrak{k}^{(N)}_0$ is strictly monotonically increasing with $N$, diverging $\sim\ln N$ to $\infty$ as $N\to\infty$.}
\smallskip

 By Theorem~2 the critical hypersurface defines a unique critical temperature $T_c(\lambda,P)$ at least for
all $\lambda >\lambda_*(P)$, and Theorem~4 in concert with Proposition~1 gives us an explicit upper bound for $\lambda_*(P)$.
\smallskip

\noindent
{\bf Corollary~1}:
 \textsl{For each $P\in \mathcal{P}$ we have the explicit upper estimate}
\begin{equation}\label{eq:lambdaSUBstarFOUR}
\lambda_*(P) <\frac{1}{\mathfrak{k}^{(4)}\big(P,T_*(P)\big)}.
\end{equation}
\smallskip

 We also obtained a rigorous lower bound on $\Lambda(P,T)$ by estimating from above the spectral radii of several operators
defining $\mathfrak{K}(P,T)$.
\smallskip

\noindent
{\bf Theorem~6}: \textsl{Let $(P,T)$ be given. Then $\mathfrak{k}(P,T)\leq \mathfrak{k}^*(P,T)$, where
\begin{align}\label{eq:kSTAR}
\hspace{-0.75truecm}
\mathfrak{k}^*(P,T)
= 
\mathfrak{k}^{(1)}(P,T) + 2\sqrt{\big(2^{1+\eps}-1\big)\zeta(1+\eps)\zeta(5-\eps)}\, \big\langle\varpi^2\big\rangle
\end{align}
with $\eps=0.65$, 
and where $\big\langle\varpi^2\big\rangle
 := \displaystyle\int_0^\infty\!\!\varpi^2 P(d\omega)$ with $\varpi={\omega}/{2\pi T}$.} 
\smallskip

\noindent
{\bf Corollary~2}:
 \textsl{By Jensen's inequality, $P(d\omega)$-averaging the concave function $\omega^2\mapsto\frac{\omega^2}{\omega^2 + 4\pi^2T^2}$ 
yields the weaker upper bound $\mathfrak{k}(P,T)\leq \mathfrak{k}^{\sharp}(P,T)$, where}
\begin{align}\label{eq:kSTARstar}
\hspace{-0.75truecm}
\mathfrak{k}^{\sharp}(P,T) = 
\tfrac{\big\langle\varpi^2\big\rangle}{\big\langle\varpi^2\big\rangle + 1}
+ 2\sqrt{\big(2^{1+\eps}-1\big)\zeta(1+\eps)\zeta(5-\eps)}\, \big\langle\varpi^2\big\rangle
\end{align}
\textsl{with $\eps=0.65$.
 Moreover, purging the $\big\langle\varpi^2\big\rangle$ contribution in the denominator of the first term at r.h.s.(\ref{eq:kSTARstar})
yields a yet weaker upper bound on $T_c(P,T)$ that is $\propto \big\langle\varpi^2\big\rangle=\big\langle\omega^2\big\rangle/4\pi^2T^2$.}
\smallskip

 We register that the explicit upper bounds (\ref{eq:kSTAR}) and (\ref{eq:kSTARstar}) 
on $\mathfrak{k}(P,T)$ are averages of functions that are
manifestly strictly monotone decreasing with $T$ on $\RR_+$, and therefore themselves strictly monotone decreasing with $T$ on $\RR_+$.

 Moreover, since $P(d\omega)$ is compactly supported, we register also that $\mathfrak{k}^*(P,T)$, and therefore also $\mathfrak{k}(P,T)$, 
are bounded above by $C/T^2$ for all $T>0$.
 Thus the largest eigenvalue of $\mathfrak{K}(P,T)$ not only overall decreases when $T$ increases, it overall decreases to 0 at least 
as fast as $C/T^2$. 
 This in itself does not imply monotonic decrease to zero, of course; 
however, together with Proposition~1, strictly monotonic decrease to zero follows for when $T>T_*(P)$.

 By Theorem~1, the critical hypersurface $\mathscr{S}_{\!c}$ in the set of $(\lambda,P,T)$ parameters
is a graph over the set $\{(P,T)\}$.
 By Theorems~3,~4, and~6 in concert, that graph is sandwiched between $1/\mathfrak{k}^*(P,T)$
(explicit lower bound) and $1/\mathfrak{k}^{(N)}(P,T)$ for any $N\in\NN$ (a decreasing sequence of upper bounds, the first four of which
are explicit).

 By the monotonicity of $T\mapsto\Lambda(P,T)$ for $T>T_*(P)$, 
the inverse function of the map $T\mapsto \Lambda(P,T)$ exists for $T>T_*(P)$.
 Thus the critical hypersurface $\mathscr{S}_{\!c}$ is
also a graph over a region of the $(\lambda,P)$ crystal parameter space where $\lambda>\lambda_*(P)$, there defining
the critical temperature $T_c(\lambda,P)$.
 From a theoretical perspective it is of course important to understand how $T_c$ depends on $\lambda$, given $P$.
 Our upper and lower bounds on $\Lambda(P,T)$ are also monotonic in $T$ and yield lower and upper bounds on $T_c(\lambda,P)$
that offer some insights into the behavior of $T_c(\lambda,P)$ itself.

 By its strict monotonic dependence on $T\in\RR_+$, in principle 
our explicit upper bound $\mathfrak{k}^*(P,T)$ on $\mathfrak{k}(P,T)$ can be inverted to yield an upper 
\textsl{critical-temperature bound} $T_c^*(\lambda,P)$ for all $\lambda>0$ and all $P\in\mathcal{P}$.
 Furthermore, our explicit lower bounds $\mathfrak{k}^{(N)}(P,T)$, $N\in\{1,2,3,4\}$, on $\mathfrak{k}(P,T)$ can in principle 
be inverted for $T>T_*(P)$ to yield lower \textsl{critical-temperature bounds} $T_c^{(N)}(\lambda,P)$ for $N\in\{1,2,3,4\}$ when 
$\lambda>\max\{\lambda_*(P),\lambda_N\}$.
 None of these bounds on $T_c$ can generally be expressed in closed form, though.

 It is worthy of note, though, that a large $T$ analysis of the operators $\mathfrak{K}^{(N)}(P,T)$ reveals that the 
large-$\lambda$ asymptotics of $T_c(\lambda,P)$ can be computed explicitly in closed form; see Corollary~3 below, which 
is a consequence of 
\smallskip

\noindent
{\bf Theorem~7}:$\!$ \textsl{The eigenvalues $\mathfrak{k}^{(N)}(P,T)$ are analytic about $T\!=\!\infty$:}
\begin{equation}\label{eq:CcurveNasympSMALL}
\hspace{-0.5truecm}
\mathfrak{k}^{(N)}(P,T)
= \tfrac{\mathfrak{g}^{(N)}(2) \left\langle\omega^2\right\rangle}{4\pi^2}\tfrac{1}{T^2}
- \tfrac{ \left\langle \mathfrak{G}^{(N)}(4)\right\rangle^{}_{\!2}\left\langle \omega^4\right\rangle }{16\pi^4}
\tfrac{1}{T^4} + \mathcal{O}\Big(\tfrac{1}{T^6}\Big);\!\!\!
\end{equation}
\textsl{here, $\mathfrak{g}^{(N)}(2)\geq 1$ is the largest eigenvalue for the $N$-Matsubara frequency approximation 
to the operator $\mathfrak{G}(\gamma)$ of the $\gamma$ model at $\gamma=2$, and 
\begin{equation}
\big\langle \mathfrak{G}^{(N)}(4)\big\rangle^{}_2 
:= \big\langle \Xi^{\mbox{\tiny{\rm opt}}}_N(2),\, \mathfrak{G}^{(N)}(4)\,\Xi^{\mbox{\tiny{\rm opt}}}_N(2) \big\rangle > 0 
\end{equation}
denotes the quantum-mechanical expected value of 
the $N$-Matsubara frequency approximation to the operator $\mathfrak{G}(\gamma)$ at $\gamma=4$,
taken with the normalized $N$-frequency  optimizer $\Xi^{\mbox{\tiny{\rm opt}}}_N(\gamma)$
of the $\gamma$  model at $\gamma=2$.
 Moreover, $\big\langle\omega^{k}\big\rangle := \int_0^\infty\!\omega^{k} P(d\omega)$ for $k\in\NN$.}

\noindent
{\bf Corollary\! 3}:\! \textsl{The $N$-Matsubara frequency approximation ${\mathscr{S}}_c^{(N)}\!\!$~to~the  critical hypersurface
${\mathscr{S}}_c$ in the positive $(\lambda,P,T)$ cone is asymptotic to a graph over the asymptotic
$\lambda\sim\infty$ region of the $(\lambda,P)$ set, given~by}
\begin{equation}\label{eq:TcNlambdaASYMP}
\hspace{-0.5truecm}
T_c^{(N)}(\lambda,P) 
\sim 
\frac{1}{\sqrt{2\pi^2\frac{\mathfrak{g}^{(N)}(2)\left\langle\omega^2\right\rangle}
{\left\langle \mathfrak{G}^{(N)}(4)\right\rangle^{}_2\left\langle\omega^4\right\rangle}
\biggl(1-\sqrt{1-4\frac{\left\langle \mathfrak{G}^{(N)}(4)\right\rangle^{}_2 \left\langle\omega^4\right\rangle} 
{\mathfrak{g}^{(N)}(2)^2\left\langle\omega^2\right\rangle^2
}{\displaystyle\frac1\lambda} }\,\biggr)}}.
\end{equation}
\textsl{This result also holds when $N\to\infty$ (with superscripts ${}^{(N)}$ purged).}
\smallskip

 Since (\ref{eq:TcNlambdaASYMP}) is valid aysmptotically for large $\lambda$,
expanding the inner square root at r.h.s.(\ref{eq:TcNlambdaASYMP}) to leading order in powers of $1/\lambda$ yields
$T_c^{(N)}(\lambda,\Omega) \sim \frac{1}{2\pi}\sqrt{\mathfrak{g}^{(N)}(2)\left\langle\omega^2\right\rangle\lambda\,}$, 
with $N\in\NN$.
 By a simple convexity estimate, 
r.h.s.(\ref{eq:TcNlambdaASYMP})$\leq \frac{1}{2\pi}\sqrt{\mathfrak{g}^{(N)}(2)\left\langle\omega^2\right\rangle\lambda\,}$.
 And so, for large enough $\lambda$ the asymptotic expression 
$\frac{1}{2\pi}\sqrt{\mathfrak{g}^{(N)}(2)\left\langle\omega^2\right\rangle\lambda\,}$
is an upper bound on $T_c^{(N)}(\lambda,P)$ that is asymptotically sharp for $\lambda\sim\infty$.
 Moreover, in \cite{KAYgamma} we showed that $\mathfrak{g}^{(N)}(2)$ converges upward to $\mathfrak{g}(2)$.
 Furthermore, as noted earlier, each $T_c^{(N)}(\lambda,P)$ vanishes for $\lambda\leq \lambda_N$, while $\sqrt{\lambda}>0$ 
for all $\lambda$. 
 All the above now suggests
\smallskip

\noindent
{\bf Conjecture~2}: \textsl{There is a critical temperature $T_c(\lambda,P) >0$ which 
for all $P\in\mathcal{P}$ and $\lambda >0$ is bounded above by $T_c(\lambda,P)< T_c^{\sim}(\lambda,P)$, with
\begin{equation}\label{eq:CONJbound}
T_c^{\sim}(\lambda,P) := \tfrac{1}{2\pi}\sqrt{\mathfrak{g}(2)\left\langle \omega^2\right\rangle\lambda\,};
\end{equation}
here, $\mathfrak{g}(2)$ is the spectral radius of $\mathfrak{G}(\gamma)$ at $\gamma=2$.}
\smallskip

 For the numerical approximation of $\frac{1}{2\pi}\sqrt{\mathfrak{g}(2)}$ to 10 significant decimal places,
see Remark~1 after Theorem~2. 

\noindent
{\bf Remark~3}: \textsl{
The asymptotic behavior 
$T_c(\lambda,P)\sim \tfrac{1}{2\pi}\sqrt{\mathfrak{g}(2)\left\langle \omega^2\right\rangle\lambda\,}$ was
first stated in \cite{AllenDynes}, see their eq.(22). 
 Apparently they did not notice that this is presumably an upper bound on $T_c$ for all $\lambda>0$.}
\smallskip

 Explicit upper and lower bounds on $T_c(\lambda,P)$ valid for all $\lambda>0$ and $P\in\mathcal{P}$ can be obtained
at the expense of less accuracy.
 For this we can use our weaker bounds on $\Lambda(P,T)$ that involve only 
$\big\langle\omega^2\big\rangle$ as $P$-average, and $\overline\Omega(P)$, both of which can be treated as numerical parameters,
and obtain the following explicit bounds on $T_c$.

\noindent
{\bf Corollary~4}:
\textsl{Whether $T_c(\lambda,P)$ is well-defined only for $\lambda>\lambda_*(P)$ or for all $\lambda>0$, given $P\in\mathcal{P}$,
it obeys the upper bound $T_c(\lambda,P)< T_c^{\sharp}(\lambda,P)$, with
\begin{equation}\label{eq:TcSTAR}
T_c^{\sharp}(\lambda,P) 
=
\tfrac{\big\langle\omega^2\big\rangle}{2\pi}\sqrt{\tfrac12\Big(\lambda(1+b) -1 + \sqrt{\big(\lambda(1+b)-1\big)^2+4b\lambda}\Big)}
\end{equation}
with $b:=2\Big(\big(2^{1+\eps}-1\big)\zeta(1+\eps)\zeta(5-\eps)\Big)^\frac12$ and $\eps=0.65$.}

  \textsl{Furthermore, for $\lambda > \overline\Omega^2/\langle\omega^2\rangle$ we have the lower bound
$T_c(\lambda,P) > T_c^{\flat}(\lambda,P)$, with}
\begin{equation}\label{eq:TcONE}
T_c^{\flat}(\lambda,P) 
=
\tfrac{1}{2\pi}\Big(\lambda\langle\omega^2\rangle -\overline\Omega^2\Big)^{\!\frac12}\!(P).
\end{equation}

 Before we next turn to the proofs of our results, we illustrate our bounds in two figures.
 The upper bound $T_c^\sharp(\lambda,P)$ on $T_c(\lambda,P)$, as
stated in Corollary~4, and the conjectured upper bound $T_c^{\sim}(\lambda,P)$ of Conjecture~2, 
and for several choices of $\overline\Omega^2/\langle\omega^2\rangle$ also the lower bound $T_c^\flat(\lambda,P)$,
are displayed in Fig.~1 as functions of $\lambda$.
 In Fig.~2 we complement Fig.~1 by
displaying the large-$\lambda$ behavior of the lower and upper approximations to $T_c(\lambda,P)$.
\newpage

\begin{figure}[H]
 \centering
\includegraphics[width=0.55\textwidth,height=0.5\textwidth,scale=0.6]{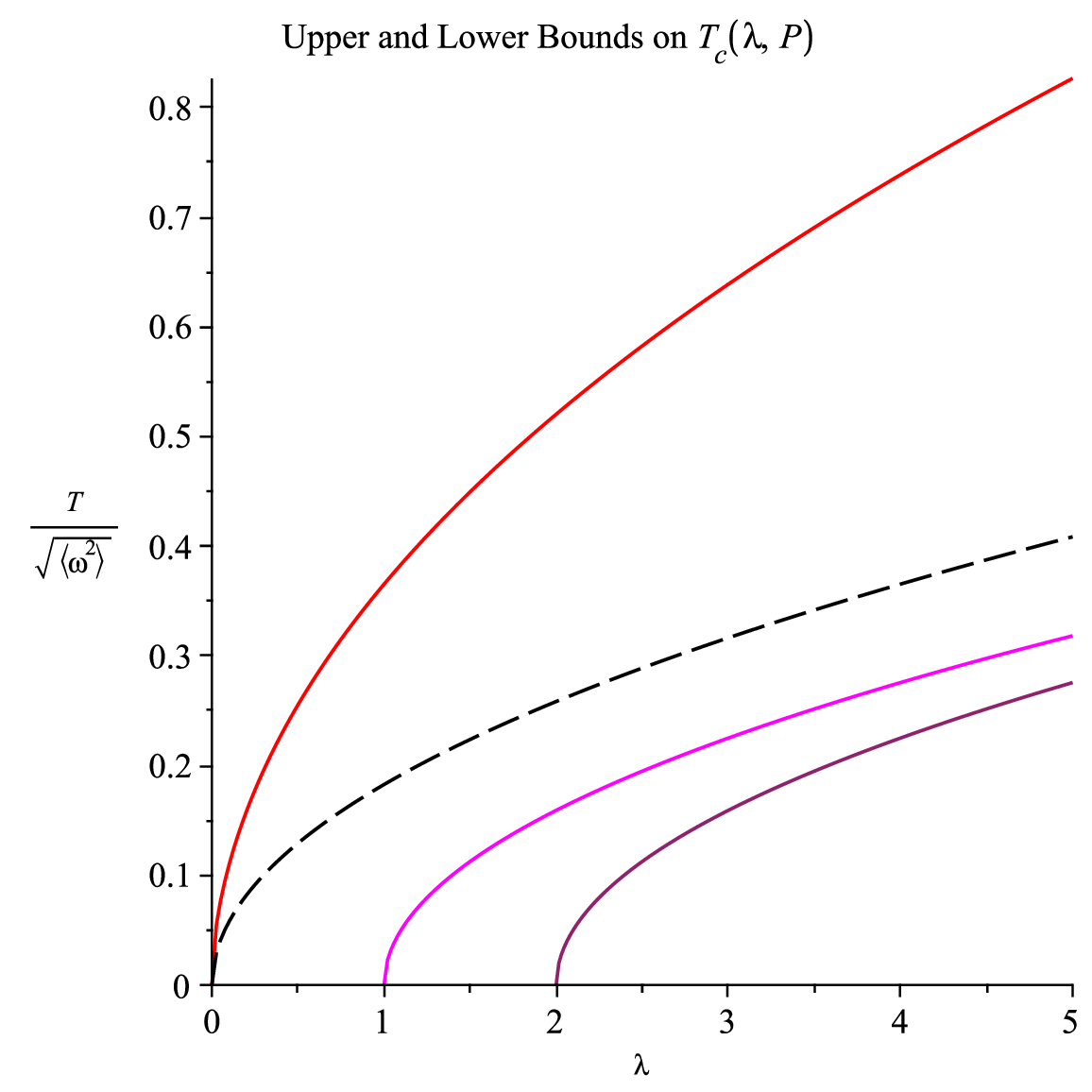} 
\vspace{-10pt}
\caption{\label{fig:TcBOUNDSvsLAMBDAdispersive}
\footnotesize
 Shown for the standard Eliashberg model are: the graph of the upper bound
$\lambda\mapsto T_c^\sharp(\lambda,P)/\sqrt{\langle\omega^2\rangle}$ (red),
the graph of the lower bound $\lambda\mapsto T_c^\flat(\lambda,P)/\sqrt{\langle\omega^2\rangle}$ for 
the values 1 (magenta) and 2 (maroon) of the ratio
$\overline\Omega^2/\langle\omega^2\rangle$, and the 
graph of the map $\lambda\mapsto T_c^{\sim}(\lambda,P)/\sqrt{\langle\omega^2\rangle}$ (black,
dash) which is asymptotic to $T_c^{}(\lambda,P)/\sqrt{\langle\omega^2\rangle}$ for $\lambda\sim\infty$.}
\end{figure}
\vspace{-15pt}

\begin{figure}[H]
 \centering
\includegraphics[height=0.5\textwidth,width=0.55\textwidth,scale=0.6]{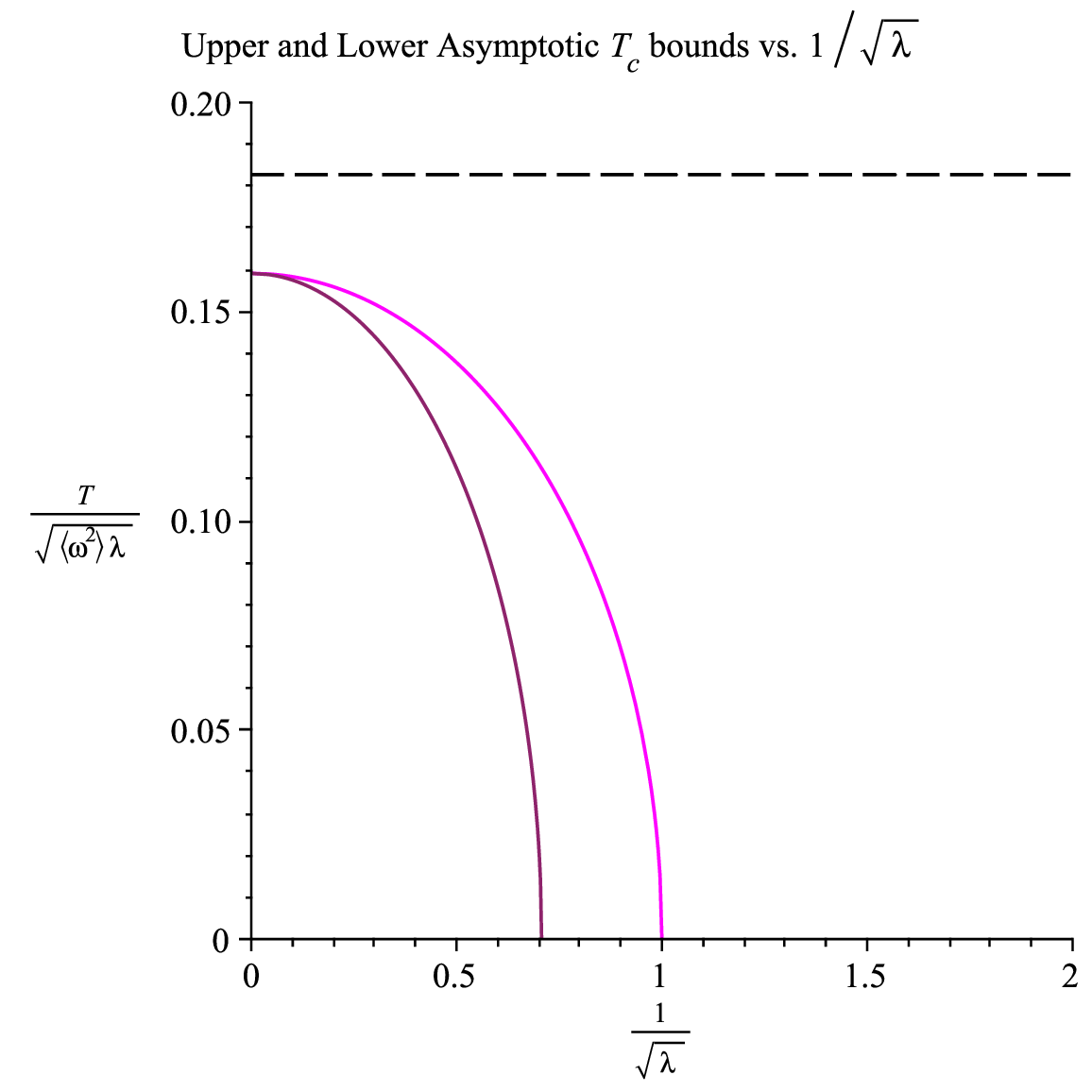} 
\caption{\label{fig:TcBOUNDSlargeLAMBDA}
\small$\!\!\!$
Shown are the graphs of the lower bound
$\sqrt{\lambda}^{-1}\! \mapsto T_c^\flat(\lambda,P)\big/\sqrt{\left\langle \omega^2\right\rangle}\sqrt{\lambda}$ for
$\overline\Omega^2/\langle\omega^2\rangle\!\in\{1,2\}$,
and the map
$\sqrt{\lambda}^{-1}\! \mapsto T_c^\sim(\lambda,P)\big/{\sqrt{\left\langle \omega^2\right\rangle}\sqrt{\lambda}}$.
 The graph of the upper bound
$\sqrt{\lambda}^{-1}\! \mapsto T_c^\sharp(\lambda,P)\big/\sqrt{\langle\omega^2\rangle}\sqrt{\lambda}$ is not shown,
for it would appear as an essentially horizontal line parallel to the black dashed line, yet a factor $\approx 2$ higher.
}
\end{figure}
\vspace{-10pt}

We now turn to the proofs of our results.
\newpage

\section{$\!\!$Stability analysis of the normal state}\label{sec:STABanal}
\label{section:stability}

 In this section we will show that for all $P\in\mathcal{P}$ and $T> 0$ there is a unique $\lambda = \Lambda(P,T)>0$
such that the normal state ${\bf N}$ is linearly stable for $\lambda<\Lambda(P,T)$, 
but unstable against superconducting perturbations for $\lambda>\Lambda(P,T)$.
 Moreover, we will show that $\Lambda(P,T)$ is characterized by a variational principle, as indicated by (\ref{eq:LambdaVP}).

 Indeed, we pave the ground precisely as in \cite{KAYgamma} by first
switching to a convenient parameterization of the Bloch spin chains.
 The symmetry relationship
 ${\bf N}_0\cdot {\bf S}_{-n} = - {\bf N}_0\cdot{\bf S}_{n-1}$  and 
{${\bf K}_0\cdot {\bf S}_{-n} =  {\bf K}_0\cdot{\bf S}_{n-1}$}, for all $n\in{\ZZ}$, allows us
to work with effective spin chains ${\bf S}\in  ({\SP}^1)^{{\NN}_0}$, with ${\NN}_0:={\NN}\cup\{0\}$.
 All summations thus go over ${\NN}_0:={\NN}\cup\{0\}$ instead of ${\ZZ}$.

 Vectors ${\bf S}_n\in\SP^1$ are represented by introducing an angle $\theta_n\in {\RR}/(2\pi{\ZZ})$ ($= [0,2\pi]$ 
with $2\pi$ and $0$ identified)
defined through ${\bf N}_0\cdot{\bf S}_n =: \cos \theta_n$ for all\footnote{If one also
introduces angles for spins with negative suffix by defining ${\bf N}_0\cdot{\bf S}_n =: \cos \theta_n$ for 
all $n\in -{\NN}$, a sequence of angles with non-negative 
suffix yields the angles with negative suffix as $\theta_{-1}=\pi-\theta_0$, $\theta_{-2}=\pi-\theta_1$, etc.,
thanks to the symmetry of ${\bf S}\in ({\SP}^1)^{{\ZZ}}$ with respect to the sign switch of the Matsubara frequencies.}
 $n\in {\NN}_0$.
 Setting $ {H}({\bf S}|{\bf N}) =: 4\pi^2 T K(\Theta)$ with $\Theta:=(\theta_n)^{}_{n\in{\NN}_0}$ yields
\begin{align}\label{eq:K}
\hspace{-1truecm}
K(\Theta) = & {\sum\limits_n} 
\big(2 n + 1\big) \big(1-\cos\theta_n \big)  \\
\notag 
\hspace{-1truecm} &
+ \frac{1}{2}
{\sum\!\sum\limits_{\hskip-0.4truecm n\neq m}} \lambda(n-m) \Big(1-\cos\big(\theta_n-\theta_m\big) \Big)
  \\
\notag 
\hspace{-1truecm} &
- \frac{1}{2}
{\sum\!\sum\limits_{\hskip-0.4truecm n\,,\, m}} 
 \lambda(n+m+1)\Big(1-\cos\big(\theta_n+\theta_m\big)\Big);
\end{align}
here, the summations run over ${\NN}_0$, and $\lambda(j)$ has been defined in (\ref{eq:VphononGENERALalt})

 The minimizing sequences of angles satisfy the {non-linear} Euler--Lagrange equation
for any stationary point $\Theta^s$ of $K(\Theta)$; viz., $\forall n\in{\NN}_0$:
\begin{align}\label{eq:EL}
\hspace{-0.8truecm}
\big( 2n + 1 \big) \sin\theta_n^s = & {\textstyle\sum\limits_{m \geq 0}} 
   \lambda(n+m+1) \sin\big(\theta_n^s+\theta_m^s\big) \\ \notag
 & - {\textstyle\sum\limits_{m \geq 0}} 
   \lambda(n-m) \sin\big(\theta_n^s-\theta_m^s\big).
\end{align}
 In the following we shall omit the superscript ${}^s$ from $\Theta^s$.

 The system of equations (\ref{eq:EL}) has infinitely many solutions when the $\theta_n$ are allowed to take
values in $[0,2\pi]$, restricted only by the asymptotic condition that $\theta_n\to 0$ rapidly enough when $n\to\infty$; 
see~\cite{YuzAltPRB}. 
 Here we are only interested in solutions that are putative minimizers of $H({\mathbf S}|{\mathbf N})$, hence of $K(\Theta)$.
 In \cite{YuzAltPRB} it was shown that a sequence $\Theta=(\theta_n)^{}_{n\in{\NN}_0}$ that minimizes 
$K(\Theta)$, must have $\Theta\in [0,\frac\pi2]^{{\NN}_0} =:S$; i.e.,
all\footnote{Alternatively, all $\theta_n\in[-\frac\pi2,0]$; these choices are gauge equivalent.}
 $\theta_n\in[0,\frac\pi2]$.

 The normal state is the sequence of angles $\underline\Theta:=(\theta_n =0)^{}_{n\in{\NN}_0}$.
 This trivial solution of (\ref{eq:EL}) manifestly exists for all $\lambda>0$ and $P$.
 However, the trivial solution $\underline{\Theta}$ representing the normal state 
minimizes $K(\Theta)$ only for sufficiently large $T$, as we will prove in a separate paper.
 In this paper we will only prove that $\underline{\Theta}$ is linearly stable against small superconducting 
perturbations $\Theta\in S$ for which $K(\Theta)$ is well-defined if and only if $T$ is large enough.

 This will prove Theorems~1 and~3.

\subsection{Proof of Theorems~1 and~3}\label{sec:THMoneANDtwoPROOFS}

 Since $K(\underline\Theta)= 0 = H(\mathbf{N}|\mathbf{N})$, to inquire into the question of linear stability versus instability 
of $\underline\Theta$ we expand $K(\Theta)$ about $\Theta=\underline\Theta$ to second order in $\Theta$.
 This yields the quadratic form
\begin{align}\label{eq:KTWOsimplified}
\hspace{-1truecm}
K^{(2)}(\Theta) = &\, 
{\sum\limits_n} \biggl[
\frac{2 n + 1}{2} 
             - \frac12 \lambda(2n+1)
+ {\sum\limits_{k=1}^{n}} \lambda(k)\biggr] \theta_n^2  \\
\notag &
\!\!- \frac{1}{2}
\sum 
\!\sum\limits_{\hspace{-0.5truecm}n\neq m} 
\theta_n\biggl[\lambda 
(n-m) + \lambda(n+m+1)\biggr]\theta_m,
\end{align}
which for all $\lambda>0$ and $P\in\mathcal{P}$ is well-defined on the Hilbert space $\mathcal{H}$ of
sequences that satisfy $\|\Theta\|_{\mathcal{H}}^2:=\sum_{n\geq 0} (2n+1)\theta_n^2 < \infty$.
 If $K^{(2)}(\Theta) \geq 0$ for all $\Theta\in\mathcal{H}$, with ``$=0$'' iff $\Theta = \underline\Theta$, 
then $K(\Theta)>0$ for all $\Theta\neq\underline\Theta$ in a sufficiently small neighborhood of
$\underline\Theta$, and then the trivial sequence $\underline\Theta$ is a local minimizer of $K(\Theta)$
and thus linearly stable.
 If on the other hand there is at least one $\Theta\neq \underline\Theta$ in $\mathcal{H}\cap S$
for which $K^{(2)}(\Theta)<0$, then the trivial sequence $\underline\Theta$ is 
not a local minimizer of $K(\Theta)$ in $\mathcal{H}\cap S$, and therefore unstable against ``superconducting perturbations.''
 The verdict as to linear stability versus instability depends on $\lambda$ and $P$.

 As in \cite{KAYgamma}, we now recast the functional $K^{(2)}(\Theta)$ defined on $\mathcal{H}$ as a functional $Q(\Xi)$ defined 
on $\ell^2({\NN}_0)$.
 For this we note that we can take the square root of the diagonal matrix $\mathfrak{O}$ whose diagonal elements are 
the odd natural numbers.
 Its square root is also a diagonal matrix, and its action on $\Theta$ componentwise is given as
\begin{equation}\label{eq:Dop}
(\mathfrak{O}^{\frac12}\Theta)_n =  \sqrt{2n + 1}\; \theta_n =: \xi_n.
\end{equation}
 Since $\Theta := (\theta_n)_{n\in\NN_0}\subset \mathcal{H}$, the sequence $\Xi:= (\xi_n)_{n\in\NN_0}\subset\ell^2(\NN_0)$.
The map $\mathfrak{O}^{\frac12}\!:\! \mathcal{H}\! \to\! \ell^2({\NN}_0)$ is invertible.
 We set $K^{(2)}(\Theta) =:\frac12 Q(\Xi)$, viz.
 \begin{align}\label{eq:QofXI}
Q(\Xi) = &\, {\sum\limits_n} \biggl[
 1 + \frac{2}{2 n + 1}
 {\sum\limits_{k=1}^{n}} \lambda(k) \biggr] \xi_n^2  \\
\notag &
- \sum\!\sum\limits\limits_{\hskip-0.5truecm n\neq m} 
\xi_n  \biggl[\frac{\lambda (n-m)}{\sqrt{2 n + 1} \, \sqrt{2 m + 1}}\biggr] \xi_m \\
\notag &
- \sum\limits_n\!\sum\limits_m     
\xi_n  \biggl[\frac{\lambda(n+m+1)}{\sqrt{2 n + 1}\, \sqrt{2 m + 1}}\biggr] \xi_m,
\end{align}
where the contributions from the first line at r.h.s.(\ref{eq:QofXI}) are positive, those from the second and third line
negative.
\smallskip

We now state the main properties of the functional $Q(\Xi)$, which decides the question of linear stability vs. 
instability of the normal state against superconducting perturbations, as a theorem. 
 Note that this theorem restates Theorem~1 in terms of the spin chain model. 
\smallskip

\noindent
{\bf Theorem~$1^+$}:
 \textsl{Let $\lambda>0$ and $P\in\mathcal{P}$ be given. Then the functional $Q$ in (\ref{eq:QofXI}) has 
a minimum on the sphere $\big\{\Xi\in\ell^2(\NN_0): \|\Xi\|_{\ell^2}=1\big\}$.
 The minimizing (optimizing) eigenmode $\Xi^{\mbox{\tiny{\rm opt}}}$ satisfies 
$\Xi^{\mbox{\tiny{\rm opt}}} \in \RR_+^{\mathbb{N}_0}$. 
  Moreover, for each $P$ and $T>0$ there is a unique $\Lambda(P,T)>0$ at which
$\min\big\{Q(\Xi) \!:\!  \|\Xi\|_{\ell^2}=1\big\} =0$, and $\min\big\{Q(\Xi) \!:\!  \|\Xi\|_{\ell^2}=1\big\} >0$ 
whenever $\lambda <\Lambda(P,T)$, while $\min\big\{Q(\Xi) \!:\!  \|\Xi\|_{\ell^2}=1\big\} <0$ when $\lambda>\Lambda(P,T)$.
 Furthermore, the map $(P,T)\mapsto\Lambda(P,T)$ is continuous in both variables.}
\smallskip

 Following closely \cite{KAYgamma}, we prepare the proof of Theorem~$1^+$ by defining several linear operators that act on 
$\ell^2(\NN_0)$ and which are associated with $Q$. 
 Letting $\big\langle \Xi,\widetilde\Xi\big\rangle$ denote the usual $\ell^2(\NN_0)$ inner product between
two $\ell^2$ sequences $\Xi$ and $\widetilde\Xi$, we write $Q$ shorter thus:
 \begin{equation}\label{eq:QofXIrew}
Q(\Xi) = 
 \bigl\langle \Xi\, , \big( \mathfrak{I} - \lambda \mathfrak{K}\big)  \Xi\bigr\rangle \,.
 \end{equation}
Here, $\mathfrak{I}$ is the identity operator, and $\mathfrak{K} = -\mathfrak{K}_1 + \mathfrak{K}_2 + \mathfrak{K}_3$, 
where the $\mathfrak{K}_j = \mathfrak{K}_j(P,T)$ for $j\in\{1,2,3\}$ are 
given by the $P$ averages $\int_0^\infty \mathfrak{H}_j(\varpi) P(d\omega)$ of the operators 
$\mathfrak{H}_j(\varpi)$ of the non-dispersive model, which componentwise act as follows: 
 \begin{align}
\label{eq:HopsCompsONE}
\hspace{-0.8truecm}
(\mathfrak{H}_1(\varpi)\Xi)_n = &\, 
 \biggl[\frac{2}{2 n + 1} {\sum\limits_{k=1}^{n}} \frac{\varpi^2}{\varpi^2 + k^2}\biggr] \xi_n \,, \\
\label{eq:HopsCompsTWO}
\hspace{-0.8truecm}
(\mathfrak{H}_2(\varpi)\Xi)_n = &\, 
\sum\limits_{m\neq n} 
\biggl[\frac{1}{\sqrt{2 n + 1}}\,\frac{\varpi^2} {\varpi^2+(n-m)^2}\,\frac{1}{\sqrt{2 m+1}}\biggr]\xi_m\,,\\
\label{eq:HopsCompsTHREE}
\hspace{-0.8truecm}
(\mathfrak{H}_3(\varpi)\Xi)_n = &\, 
\sum\limits_m \biggl[\frac{1}{\sqrt{2 n + 1}}\, \frac{\varpi^2}{ \varpi^2 +(n+m+1)^2}\, \frac{1}{\sqrt{2 m + 1}}\biggr]\xi_m\,.
\end{align}
 Note that $\mathfrak{H}_1$ is a diagonal operator with non-negative diagonal elements, 
$\mathfrak{H}_2$ is a real symmetric operator with vanishing diagonal elements and positive off-diagonal elements,
and $\mathfrak{H}_3$ is a real symmetric operator with all positive elements. 
 Therefore, also $\mathfrak{K}_1$ is a diagonal operator with non-negative diagonal elements, 
$\mathfrak{K}_2$ is a real symmetric operator with vanishing diagonal elements and positive off-diagonal elements,
and $\mathfrak{K}_3$ is a real symmetric operator with all positive elements. 
 Note furthermore that the operators $\mathfrak{K}_j$ do not dependend on the parameter $\lambda=\lambda(0)$.

 Next, analogous to \cite{KAYgamma} we have the following result.
\smallskip

\noindent
{\bf Proposition~2}: 
\textit{For $j\in\{1,2,3\}$, each $\mathfrak{K}_j\in \ell^2(\NN_0\times\NN_0)$ for all $T > 0$ and $P\in\mathcal{P}$.
 Thus the operators $\mathfrak{K}_j = \mathfrak{K}_j(P,T)$ for $j\in\{1,2,3\}$ are Hilbert--Schmidt 
operators that map $\ell^2(\NN_0)$ compactly into $\ell^2(\NN_0)$.}
\smallskip

\noindent
\textsl{Proof of Proposition~2}: 
 We recall that $\lim_{\varpi\to 0} \frac{1}{\varpi^2}\mathfrak{H}_j(\varpi)= \mathfrak{G}_j(2)$ for
$j\in\{1,2,3\}$, with $\mathfrak{G}(\gamma) = -\mathfrak{G}_1(\gamma) + \mathfrak{G}_2(\gamma) +\mathfrak{G}_3(\gamma)$
the interaction matrix of the linearized $\gamma$ model.
 In Appendix~A of \cite{KAYgamma} we proved that each $\mathfrak{G}_j(\gamma)$, $j\in\{1,2,3\}$ is a Hilbert--Schmidt operator for
all $\gamma>0$. 
 In particular, the $\mathfrak{G}_j(2)$, $j\in\{1,2,3\}$ are Hilbert--Schmidt operators.
 We let $\big[{\mathfrak{H}_j(\varpi)}\big]_{n,m}$ denote the expressions bracketed $\big[\cdots\big]$
in (\ref{eq:HopsCompsONE}), (\ref{eq:HopsCompsTWO}), (\ref{eq:HopsCompsTHREE}).
 Since the maps $\varpi\mapsto \frac{1}{\varpi^2}\big[{\mathfrak{H}_j(\varpi)}\big]_{n,m}$ 
are non-negative and strictly monotonically decreasing for each $j\in\{1,2,3\}$, it follows that 
also the operators $\frac{1}{\varpi^2}\mathfrak{H}_j(\varpi)$ are Hilbert--Schmidt operators, 
hence so are the operators $\mathfrak{H}_j(\varpi)$ for all $j\in\{1,2,3\}$.

 A moment of reflection reveals that this now implies that the $\ell^2$ norm of the operators $\mathfrak{K}_j$ is bounded above by 
$\int_0^\infty\varpi^2 P(d\omega)$ times the $\ell^2$ norm of the $\gamma$ model operators $\mathfrak{G}_j$ at $\gamma=2$. 
\hfill {\bf Q.E.D.}
\smallskip

We are ready to prove Theorem~1$^+$.

\newpage

\noindent
\textsl{Proof of Theorem~$1^+$}:  
Having Proposition~2, the proof of our Theorem~1$^+$ is a straightforward adaptation of the proof of Theorem~$1$ in \cite{KAYgamma}.
 All that needs to be changed is the following: 

 Replace the operators $\mathfrak{G}_j(\gamma)$, $j\in\{1,2,3\}$ by operators $\mathfrak{K}_j(P,T)$, $j\in\{1,2,3\}$, 
and replace $\mathfrak{G}(\gamma)$ by $\mathfrak{K}(P,T)$.

 Similarly, replace the eigenvalues $\mathfrak{g}_j(\gamma)$, $j\in\{1,2,3\}$ by the eigenvalues $\mathfrak{k}_j(P,T)$, $j\in\{1,2,3\}$, 
and replace $\mathfrak{g}(\gamma)$ by $\mathfrak{k}(P,T)$, and $\beta^\gamma$ by $\lambda$.
 
 Here it shall suffice to point out that (\ref{eq:QofXIrew}) makes it plain that 
$\min\big\{Q(\Xi) \!:\!  \|\Xi\|_{\ell^2}=1\big\} = 0$ if and only if $\lambda\mathfrak{k}(P,T)=1$,
with $\mathfrak{k}(P,T)>0$ denoting the largest eigenvalue of $\mathfrak{K}(P,T)$.
 Precisely when $\lambda = \Lambda(P,T)$, with
 \begin{equation}\label{eq:LambdaVPk}
\Lambda(P,T) =  \tfrac{1}{\mathfrak{k}(P,T)},
\end{equation}
then the pertinent eigenvalue problem for the minimizing mode $\Xi^{\mbox{\tiny{\rm opt}}}$ of $Q(\Xi)$ reads
\begin{equation}\label{eq:EVeqXiOPT}
\big( \mathfrak{I} -\Lambda(P,T)\mathfrak{K}\big) \Xi^{\mbox{\tiny{\rm opt}}} = 0,
\end{equation}
which, since $\mathfrak{k} = 1/\Lambda(P,T)$, is equivalent to 
\begin{equation}\label{eq:EVfixPTeqXiOPT}
\mathfrak{C}\big(\mathfrak{k}(P,T)\big)\Xi^{\mbox{\tiny{\rm opt}}} = \Xi^{\mbox{\tiny{\rm opt}}}
\end{equation}
where here
\begin{equation}\label{eq:Cop}
\mathfrak{C}\big(\eta\big) := 
\big(\eta\mathfrak{I} +\mathfrak{K}_1\big)^{-1} \big(\mathfrak{K}_2+ \mathfrak{K}_3\big).
\end{equation}
 As in the proof of Theorem~1 in \cite{KAYgamma} one shows that $\mathfrak{C}(\eta)$ for $\eta>0$ is
a compact operator that maps the positive cone $\ell^2_{\geq 0}(\NN_0)$ into itself, in fact mapping any non-zero element of
$\ell^2_{\geq 0}(\NN_0)$ into the interior of $\ell^2_{\geq 0}(\NN_0)$, and that
the spectral radius of $\mathfrak{C}\big(\mathfrak{h}\big)$ equals 1.
 Thus the Krein--Rutman theorem applies and guarantees that the non-trivial 
solution $\Xi^{\mbox{\tiny{\rm opt}}}$ of (\ref{eq:EVfixPTeqXiOPT}) is in the 
positive cone $\ell^2_{\geq 0}(\NN_0)$ (after at most choosing the overall sign), 
hence a superconducting perturbation of the normal state $\underline{\Xi}$.
\hfill {\textbf{Q.E.D.}}
\smallskip

 The proof of Theorem~$1^+$ also proves our Theorem~1. \hfill {\textbf{Q.E.D.}}
\smallskip

 The proof of Theorem~$1^+$ also proves our Theorem~3. \hfill {\textbf{Q.E.D.}}
\smallskip

 As in \cite{KAYgamma}, it is useful to add the following non-obvious fact about the spectrum of the operator $\mathfrak{K}(P,T)$.

\smallskip
\noindent
{\bf Proposition~3}: 
\textsl{Let $T> 0$ and $P\in\mathcal{P}$ be given. Then the largest eigenvalue $\mathfrak{k}(P,T)$ of
$\mathfrak{K}(P,T)$ is also the spectral radius $\rho\big(\mathfrak{K}(P,T)\big)$.}

\smallskip
\noindent
\textsl{Proof of Proposition~3}: 
 The proof is a straightforward adaptation from the proof of Proposition~1 in \cite{KAYgamma}, 
with the same replacements needed as stated in the proof of our Theorem~$1^+$. \hfill{\bf Q.E.D.}
\smallskip

 Proposition~3 allows us to characterize $\Lambda(P,T)$ as follows:
\begin{equation}\label{eq:LambdaVPrho}
\Lambda(P,T) =  \tfrac{1}{\rho(\mathfrak{K})(P,T)}.
\end{equation}
 Each of (\ref{eq:LambdaVPrho}) and (\ref{eq:LambdaVPk}) offer their own advantages to estimate $\Lambda$.

\section{Upper bounds on~$\Lambda(P,T)$}\label{sec:upperLambdaB}

\subsection{\hspace{-10pt}Upper bounds $\Lambda^{(N)}(P,T)$, $N\in\{1,2,3,4\}$}

 We here prove Theorem~4.
\smallskip
 
\noindent
\textsl{Proof}:
 Theorem~$1^+$, or rather its proof established that
$\Lambda(P,T) = \frac{1}{\mathfrak{k} (P,T)}$,
with $\mathfrak{k}(P,T)>0$ the largest eigenvalue of $\mathfrak{K}(P,T)$.
 More explicitly,
\begin{equation}\label{eq:LAMBDAcVP}
\Lambda(P,T) : = \frac{1}{\max_\Xi \frac{\big\langle\Xi\,,\,\mathfrak{K}(P,T) \,\Xi\big\rangle}
        {\big\langle\Xi\,,\,\Xi\big\rangle}},
\end{equation}
where the maximum is taken over non-vanishing $\Xi\in\ell^2(\NN_0)$.

 Since $\mathfrak{K} (P,T)$ is compact, in principle one can get
arbitrarily accurate upper approximations to $\Lambda(P,T)$ by restricting 
$\mathfrak{K}(P,T)$ to suitably chosen finite-dimensional subspaces of $\ell^2(\NN_0)$.
 A sequence of decreasing rigorous upper bounds on $\Lambda(P,T)$ that converges to $\Lambda(P,T)$ is
 obtained by restricting the variational principle to a sequence of subspaces of $\ell^2(\NN_0)$ consisting 
of vectors $\Xi_N:= (\xi_0,\dots,\xi_{N-1},0,\dots)$, with $\xi_j>0$ for $j\in\{0,...,N-1\}$ and $N\in\NN$.
 Evaluating (\ref{eq:LAMBDAcVP}) with $\Xi_N$ in place of 
$\Xi^{\mbox{\tiny{\rm opt}}}$ 
yields a strictly monotonically decreasing
sequence of upper bounds $\Lambda^{(N)}(P,T)$ on $\Lambda(P,T)$, viz.
\begin{equation}\label{eq:LAMBDAcVPtrialN}
\Lambda^{(N)}(P,T) : = 
\frac{1}{\max_{\Xi_N} \frac{\big\langle\Xi_N\,,\,\mathfrak{K}(P,T)\,\Xi_N\big\rangle}
        {\big\langle\Xi_N,\,\Xi_N\big\rangle}}.
\end{equation}

 The evaluation of (\ref{eq:LAMBDAcVPtrialN}) is equivalent to finding the largest eigenvalue of
a real symmetric matrix $N\times N$ matrix $\mathfrak{M}$, i.e. the largest zero of the associated
degree-$N$ characteristic polynomial of $\mathfrak{M}$.
 As noted in \cite{KAYgamma}, the coefficients $c_k$ of the characteristic polynomial 
$\det\big(\mu\mathcal{I}-\mathfrak{M}\big) =: \sum_{k=0}^N c_k\mu^k$ 
are explicitly known polynomials of degree $N-k$ in $\mathrm{tr}\, \mathfrak{M}^j$, $j\in\{1,...,N\}$.
 When $N\in\{1,2,3,4\}$ the zeros of the characteristic polynomial can be computed algebraically in closed form.
 For general real symmetric $N\times N$ matrices $\mathfrak{M}$ these spectral formulas have been listed in
\cite{KAYgamma} and need not be repeated here. 

 The task that remains is to substitute $\mathfrak{K}^{(N)}$, $N\in\{1,2,3,4\}$, for $\mathfrak{M}$ and to
select the largest eigenvalue for each $N$ from these spectra.
 This is trivial for $N=1$, yielding (\ref{eq:kONE}), and it is straightforward for $N=2$ and $N=4$,
yielding (\ref{eq:kTWO}) and (\ref{eq:kFOUR}), respectively.
 The only case that is not quite so straightforward is $N=3$. 
 In this case we argue as follows.

 We recall that $\lim_{\varpi\to 0} \frac{1}{\varpi^2}\mathfrak{H}^{(N)}(\varpi)= \mathfrak{G}^{(N)}(2)$,
the interaction matrix of the linearized $\gamma$ model at $\gamma=2$, truncated to $N$ Matsubara frequencies. 
 In \cite{KAYgamma} we determined the largest eigenvalue of $\mathfrak{G}^{(3)}(2)$, which equals its
spectral radius, then used a continuity argument, combined with the Perron--Frobenius theorem that establishes
the non-degeneracy of the spectral radius, to extend the determination of the largest eigenvalue of $\mathfrak{G}^{(3)}(2)$
to that of $\mathfrak{G}^{(3)}(\gamma)$ for all $\gamma>0$.
 By the same type of reasoning we now can extend the determination of the largest eigenvalue of $\mathfrak{G}^{(3)}(2)$
to that of $\frac{1}{\varpi^2}\mathfrak{H}^{(3)}(\varpi)$ for all $\varpi>0$.
 Multiplication by $\varpi^2$ for $\varpi>0$ does not change the ordering of the eigenvalues, and this yields
(\ref{eq:kTHREE}) for when $P$ is a Dirac delta measure, i.e. for the non-dispersive model.

 At last, since the interaction kernels $\lambda(n)$ are bounded linear functionals of $P$, they are continuous in $P$, 
and manifestly they are continuous in $T>0$.
 Therefore by the same type of reasoning we now can extend the determination of the largest eigenvalue of 
$\mathfrak{H}^{(3)}(\varpi)$ to that of the largest eigenvalue of  $\mathfrak{K}^{(3)}(P,T)$, yielding (\ref{eq:kTHREE}).

This proves Theorem~4. \hfill {\bf Q.E.D.}

 For $N>4$ a numerical approximation of $\mathfrak{k}^{(N)}(P,T)$ is necessary for any choices of $P$ and $T$ that are of interest.

\subsection{\hspace{-11pt}Upper bounds $\Lambda^{(N)}(P,T)$ as $T\to 0$, $N\!\in\!\NN$}

 We here prove Theorem~5 by evaluating $\Lambda^{(N)}(P,T)$ in the limit $T\to 0$, for all $N\in\NN$.
\smallskip
 
\noindent
\textsl{Proof of Theorem~5}: Note that $T\to 0$ means $\varpi\to\infty$ for all $\omega>0$.
 For each $N\in\NN$ one easily finds
\begin{equation}\label{eq:HNasympINF}
\lim_{\varpi\to\infty} \mathfrak{H}^{(N)}(\varpi) = - \mathfrak{I}^{(N)} + 2\, \Xi_N^*\otimes\Xi_N^*,
\end{equation}
with $\mathfrak{I}^{(N)}$ the $N\times N$ identity matrix, and where 
$\Xi_N^*= (\xi_0^*,\xi_1^*,...,\xi_{N-1}^*)$ has components $\xi_n^* = \frac{1}{\sqrt{2n+1}}$.
 Since r.h.s.(\ref{eq:HNasympINF}) is independent of $\omega$, also
\begin{equation}\label{eq:KNasympINF}
\lim_{T\to 0} \mathfrak{K}^{(N)}(P,T) = - \mathfrak{I}^{(N)} + 2\, \Xi_N^*\otimes\Xi_N^*,
\end{equation}
independently of $P$.
 The limiting matrix at r.h.s.(\ref{eq:KNasympINF}) has an $N-1$-dimensional eigenspace for the eigenvalue $-1$, 
consisting of the orthogonal complement of $\Xi_N^*$.
 The remaining eigenspace is $\{t\Xi_N^*;t\in\RR\}$, associated with the eigenvalue 
$-1+2{\textstyle\sum\limits_{n=0}^{N-1}\frac{1}{2n+1}}\geq 1$. 
 Manifestly this eigenvalue is the largest eigenvalue of the matrix at r.h.s.(\ref{eq:KNasympINF}); it also is its spectral radius.
 \hfill {\textbf{Q.E.D.}}
\smallskip

 We note that r.h.s.(\ref{eq:lambdaN}) diverges to $\infty$ when $N\to\infty$, essentially like $\ln N$. 
 Thus, $\lim_{T\to 0}\Lambda^{(N)}(P,T)=\frac{1}{\mathfrak{k}^{(N)}_0}=: \lambda_N^{} \to 0$ as $N\to \infty$, 
as claimed in the introduction.
\smallskip

\subsection{\hspace{-11pt}Upper bounds $\Lambda^{(N)}(P,T)$ at $T\!\gg\! \overline\Omega(P)$, $N\!\in\!\NN$}

We here prove Theorem~7.

\noindent
\textsl{Proof of Theorem~7}: Since $P$ is supported on $[0,\overline\Omega(P)]$,
the parameter regime $T\gg \overline\Omega(P)$ implies that $\varpi\ll 1$ uniformly in all integrals w.r.t. $P$.
 Since all kernels of $\mathfrak{H}_j(\varpi)$, $j\in\{1,2,3\}$, are of the type $\frac{\varpi^2}{\varpi^2+k^2}>0$, with
$k\in\NN$, the operators $\mathfrak{H}_j(\varpi)$, $j\in\{1,2,3\}$, and their $N$-frequency truncations,
are real analytic in $\varpi$, and in $\varpi^2$, about $\varpi =0 = \varpi^2$.
 Maclaurin expansion in powers of $\varpi^2$ about $\varpi^2=0$ yields
\begin{equation}
\frac{\varpi^2}{\varpi^2+k^2} = \varpi^2\frac{1}{k^2} - \varpi^4\frac{1}{k^4} \pm \cdots,
\end{equation}
which implies that
\begin{equation}
\mathfrak{H}_j(\varpi)= \varpi^2 \mathfrak{G}_j(2) - \varpi^4\mathfrak{G}_j(4) \pm \cdots,\quad j\in\{1,2,3\},
\end{equation}
and the analog holds for their $N$-frequency truncations.
 Averaging over $P$ then yields the asymptotic large $T$ expansion
\begin{equation}
\hspace{-0.5truecm}
\mathfrak{K}_j(P,T)= \frac{\mathfrak{G}_j(2) \langle\omega^2\rangle}{4\pi^2}\frac{1}{T^2}
   - \frac{\mathfrak{G}_j(4) \langle\omega^4\rangle}{16\pi^4}\frac{1}{T^4} \pm \cdots,\; j\in\{1,2,3\},
\end{equation}
and analogously for their $N$-frequency truncations.

 The claim (\ref{eq:CcurveNasympSMALL}) of the theorem now follows from first-order perturbation theory \cite{Kato}. 

It remains to show that $\forall\,N\in\NN$ we have 
{$\langle \mathfrak{G}^{(N)}(4)\rangle^{}_2 > 0$}, where
\begin{equation}\label{eq:expectKfourWITHtwo}
\langle \mathfrak{G}^{(N)}(4)\rangle^{}_2
 := \frac{\big\langle\Xi^{\mbox{\tiny{\rm opt}}}_N(2),\mathfrak{G}^{(N)}(4)\,\Xi^{\mbox{\tiny{\rm opt}}}_N(2)\big\rangle}
        {\big\langle\Xi^{\mbox{\tiny{\rm opt}}}_N(2),\Xi^{\mbox{\tiny{\rm opt}}}_N(2)\big\rangle} ;
\end{equation}
here, $\Xi^{\mbox{\tiny{\rm opt}}}_N(2)$ is a not necessarily normalized positive
eigenvector for the top eigenvalue $\mathfrak{g}^{(N)}(2)$ of $\mathfrak{G}^{(N)}(2)$.
 
We prove a stronger result that implies that {$\langle \mathfrak{G}^{(N)}(4)\rangle^{}_2>0$.}
\smallskip

\noindent
{\bf Proposition~4}: \textsl{Let $\gamma > 0$ be given. Then for all $\gamma^\prime >0$ and $N\in\NN_0$,}
\begin{equation}\label{eq:expectKgammaWITHgammaNULL}
\langle \mathfrak{G}^{(N)}(\gamma^\prime)\rangle^{}_{\gamma} := 
\frac{\big\langle\Xi^{\mbox{\tiny{\rm opt}}}_N(\gamma),\mathfrak{G}^{(N)}(\gamma^\prime)\,\Xi^{\mbox{\tiny{\rm opt}}}_N(\gamma)\big\rangle}
      {\big\langle\Xi^{\mbox{\tiny{\rm opt}}}_N(\gamma),\Xi^{\mbox{\tiny{\rm opt}}}_N(\gamma)\big\rangle} {> 0},
\end{equation}
\textsl{with}
$\Xi^{\mbox{\tiny{\rm opt}}}_N(\gamma)$ \textsl{any eigenvector of the
top eigenvalue $\mathfrak{g}^{(N)}(\gamma)$ of $\mathfrak{G}^{(N)}(\gamma)$.}
\smallskip

\noindent
{\it Proof of Proposition~4}: We begin by noting that {the proof is trivial when $N=1$. 
 It remains to prove the proposition for when $N>1$.}

 In \cite{KAYgamma} we already proved that
$\langle \mathfrak{G}^{(N)}(\gamma^{})\rangle^{}_{\gamma^{}} = \mathfrak{g}^{(N)}(\gamma)\geq 1$,
{with ``$=$'' holding iff $N=1$.}
 Thus we see that Proposition~4 is true when restricted to $\gamma=\gamma^\prime$.
 To see that Proposition~4 is true also when $\gamma\neq \gamma^\prime$, we need the input of another result.
\smallskip

\noindent
{\bf Proposition~5}: \textsl{Let $\gamma> 0$ be given. 
Then for all $N\in\NN_0$, the map} $n\mapsto \frac{1}{\sqrt{2n+1}}\big(\Xi^{\mbox{\tiny{\rm opt}}}_N(\gamma)\big)_n$ 
\textsl{is positive and decreasing.}
\smallskip

\noindent
{\it Proof of Proposition~5}: {Also the proof of Proposition~5 is trivial when $N=1$. 
 Henceforth $N>1$, therefore.}

{We first prove the positivity.}
The eigenvector $\Xi^{\mbox{\tiny{\rm opt}}}_N(\gamma)$ of the largest eigenvalue $\mathfrak{g}^{(N)}(\gamma)$
of $\mathfrak{G}^{(N)}(\gamma)$ solves the linear eigenvalue equation
\begin{align}
\label{eq:EVEQgamma}
\mathfrak{G}^{(N)}(\gamma)
 \Xi = \mathfrak{g}^{(N)}(\gamma)\Xi,
\end{align}
where
\begin{align}
\label{eq:Kops}
\mathfrak{G}^{(N)}(\gamma) 
= - \mathfrak{G}^{(N)}_1(\gamma) + \mathfrak{G}^{(N)}_2(\gamma) + \mathfrak{G}^{(N)}_3(\gamma)
\end{align}
is the projection of 
$\mathfrak{G}(\gamma)= - \mathfrak{G}_1(\gamma) + \mathfrak{G}_2(\gamma) + \mathfrak{G}_3(\gamma)$
onto the subspace spanned by the first $N$ positive Matsubara frequencies.
 The operators $\mathfrak{G}_j(\gamma)$ act componentwise as follows (cf. \cite{KAYgamma}),
\begin{align}
\label{eq:Kops1}
\hspace{-0.8truecm}
(\mathfrak{G}_1(\gamma)\Xi)_n = &\, 
 \biggl[\frac{1}{2 n + 1} {\sum\limits_{k=1}^{n}} \frac{2}{k^\gamma}\biggr] \xi_n \,, \\
\label{eq:Kops2}
\hspace{-0.8truecm}
(\mathfrak{G}_2(\gamma)\Xi)_n = &\, 
\sum\limits_m \biggl[\frac{1}{\sqrt{2 n + 1}}\,\frac{1-\delta_{n,m}}{|n-m|^\gamma}\,\frac{1}{\sqrt{2 m+1}}\biggr]\xi_m\,,\\
\label{eq:Kops3}
\hspace{-0.8truecm}
(\mathfrak{G}_3(\gamma)\Xi)_n = &\, 
\sum\limits_m \biggl[\frac{1}{\sqrt{2 n + 1}}\, \frac{1}{ (n+m+1)^\gamma}\, \frac{1}{\sqrt{2 m + 1}}\biggr]\xi_m\,,
\end{align}
where it is understood that $\frac{1 - \delta_{n,m}}{ |n-m|^\gamma} :=0$ when $n=m$.
 Analogously their $N$-Matsubara frequency truncations are defined by limiting $n$ and $m$ to the set $\{0,..., N-1\}$ 
in (\ref{eq:Kops1}), (\ref{eq:Kops2}), and (\ref{eq:Kops3}).
 
 Analogously to (\ref{eq:EVfixPTeqXiOPT}), we can rewrite (\ref{eq:EVEQgamma}) as the fixed point problem\begin{equation}
\label{eq:EVfixPTeqXiOPTgamma}
\mathfrak{C}_\gamma^{(N)}\big[\mathfrak{g}(\gamma)\big]\Xi_N = \Xi_N
\end{equation}
where $\mathfrak{C}_\gamma^{(N)}\big[\eta\big]$, with $\eta>0$, is the restriction of
\begin{equation}
\label{eq:CopGAMMA}
\mathfrak{C}_\gamma\big[\eta\big] := 
\big(\eta\mathfrak{I} +\mathfrak{G}_1(\gamma)\big)^{-1} \big(\mathfrak{G}_2(\gamma)+ \mathfrak{G}_3(\gamma)\big)
\end{equation}
to the subspace of $\ell^2(\NN_0)$ spanned by the first $N$ positive Matsubara frequencies.
 The subscript ${}_\gamma$ is meant as a reminder to distinguish the operator $\mathfrak{C}_\gamma\big[\eta\big]$ of the 
$\gamma$ model from the operator $\mathfrak{C}\big[\eta\big]$ of the standard Eliashberg model that we defined in (\ref{eq:Cop}).
 As shown in \cite{KAYgamma}, and analogously in the proof of Theorem~$1^+$ in this paper, the operator 
$\mathfrak{C}_\gamma\big[\eta\big]$ maps $\ell^2_{\geq0}(\NN_0)$ compactly into itself.
 Any non-zero element of $\ell^2_{\geq0}(\NN_0)$ is mapped into the interior of $\ell^2_{\geq0}(\NN_0)$.
 Iff $\eta=\mathfrak{g}(\gamma)$ its spectral radius equals $1$.
 It is non-degenerate and the associated normalized 
eigenmode $\Xi^{\mbox{\tiny{\rm opt}}}(\gamma)\in \ell^2_{\geq0}(\NN_0)$ has only positive components, by the Krein--Rutman
theorem.
  The analogous statement holds for all its finite-rank, $N$-frequency approximations
$\Xi^{\mbox{\tiny{\rm opt}}}_N(\gamma)$, by the Perron--Frobenius theorem.

Having established the positivity of $\Xi^{\mbox{\tiny{\rm opt}}}_N(\gamma)$ for all $N\in\NN$, the positivity of the
map $n\mapsto \frac{1}{\sqrt{2n+1}}\big(\Xi^{\mbox{\tiny{\rm opt}}}_N(\gamma)\big)_n$ follows trivially.

 We turn to the decrease of the map $n\mapsto \frac{1}{\sqrt{2n+1}}\big(\Xi^{\mbox{\tiny{\rm opt}}}_N(\gamma)\big)_n$. 
 Recall that 
$\Xi^{\mbox{\tiny{\rm opt}}}_N(\gamma)$ is a positive maximizer of 
$\big\langle\Xi_N\, ,\mathfrak{G}^{(N)}(\gamma)\,\Xi_N\big\rangle/\big\langle\Xi_N\, ,\,\Xi_N\big\rangle$ on $\RR^N\backslash\{0_N\}$,
where $0_N$ is the vanishing vector.
 Now recall that 
$\frac{1}{\sqrt{2n+1}}\big(\Xi_N(\gamma)\big)_n=\big(\Theta_N(\gamma)\big)_n$,
where by $\Theta_N$ we denote a truncation to the first $N$ positive Matsubara frequencies of the sequence $\Theta$ of angles 
$(\theta_n)_{n\in\NN_0}$ defined in section~\ref{sec:STABanal}, except that for the linearized problem the restriction of the 
$\theta_n$ to $[0,2\pi]$ can be dropped.
{Defining ${\mathfrak{D}}$ to be the diagonal matrix whose $n$-th diagonal entry is the $n$-th odd positive integer,
we can more compactly write $\Xi = \sqrt{\mathfrak{D}}\Theta$ and $\Xi_N = \sqrt{\mathfrak{D}}^{(N)}\Theta_N$.
 Thus we have
\begin{equation}\label{eq:XiMAXvsThetaMAXtruncated}
\max_{\RR^N\backslash\{0_N\}}
 \frac{\big\langle\Xi_N,\mathfrak{G}^{(N)}(\gamma)\,\Xi_N\big\rangle}
        {\big\langle\Xi_N,\Xi_N\big\rangle} 
\equiv \max_{\RR^N\backslash\{0_N\}} \frac{\big\langle\Theta_N,\widehat{\mathfrak{G}}^{(N)}(\gamma)\,\Theta_N\big\rangle}
        {\big\langle\Theta_N,\mathfrak{D}^{(N)}\Theta_N\big\rangle} ,
\end{equation}
where $\widehat{\mathfrak{G}}^{(N)}(\gamma)$ is the truncation to the first $N$ Matsubara frequencies of 
$\widehat{\mathfrak{G}}(\gamma):= \sqrt{\mathfrak{D}} {\mathfrak{G}}(\gamma)\sqrt{\mathfrak{D}}$ .
 Written explictly in terms of the $\theta_n$, the quadratic forms at r.h.s.(\ref{eq:XiMAXvsThetaMAXtruncated}) read
\begin{equation}\label{eq:diagFORM}
{\big\langle\Theta,\mathfrak{D}\Theta\big\rangle} := 
{\sum\limits_n} (2n+1) \theta_n^2
\end{equation}
and}
\begin{align}\label{eq:WIDEhatG}
\hspace{-1truecm}
& 
\big\langle\Theta,\widehat{\mathfrak{G}}(\gamma)\,\Theta\big\rangle
 := \\ \notag &
- {\sum\limits_n} \biggl( {\sum\limits_{k=1}^{n}} \frac{2}{k^\gamma}
\!\biggr)  \theta_n^2   + \sum\limits_n \!\sum\limits_m
\theta_n\biggl[\frac{1 - \delta_{n,m}}{ |n-m|^\gamma} + \frac{1}{(n+m+1)^\gamma}\biggr]\theta_m,
\end{align}
where $\sum_n (\cdots)$ means summation over $\NN_0$, and
where again it is understood that $\frac{1 - \delta_{n,m}}{ |n-m|^\gamma} :=0$ when $n=m$.
 The $N$-Matsubara frequency truncations of (\ref{eq:WIDEhatG}) and (\ref{eq:diagFORM}) 
are defined by limiting $n$ and $m$ to the set $\{0,..., N-1\}$.
We have completed our preparations for showing that any positive maximizer $\Theta^{\mbox{\tiny{\rm opt}}}_N(\gamma)$
of r.h.s.(\ref{eq:XiMAXvsThetaMAXtruncated}) is decreasing in $n$. 

{We will show that with the help of decreasing rearrangement ${}^\sharp\Theta_N$ 
(by suitable permutation of the entries of $\Theta_N$) we can increase 
the ratio of the two quadratic forms at r.h.s.(\ref{eq:XiMAXvsThetaMAXtruncated})
 if $\Theta_N$ was not already decreasing everywhere.
 While decreasing rearrangements generally do not preserve the normalization, this is not a problem since 
the ratio of the two quadratic forms at r.h.s.(\ref{eq:XiMAXvsThetaMAXtruncated}) is homogeneous of degree zero under scaling
of $\Theta_N$, so that after a decreasing rearrangement the stipulated normalization can be restored with a simple overall scaling,
without changing the value of the ratio of the two quadratic forms.}

 First, since the map $n\mapsto {\sum\limits_{k=1}^{n}} \frac{1}{k^\gamma}$ is strictly increasing on $\NN_0$, 
given any $\Theta_N\in\RR^N_+$ that is locally strictly increasing somewhere, 
its decreasing rearrangement ${}^\sharp\Theta_N$ obeys the inequality
${\sum\limits_n} \Big( {\sum\limits_{k=1}^{n}}\frac{1}{k^\gamma}\Big) {}^\sharp\theta_n^2  <
 {\sum\limits_n} \Big( {\sum\limits_{k=1}^{n}} \frac{1}{k^\gamma}\Big) \theta_n^2$, where $n$ here runs from $0$ to $N-1$.

Second, we note that 
\begin{align}\label{eq:RIESZfnctl}
\hspace{-1truecm}
\sum\limits_{ n} \!\sum\limits_{ m}
\theta_n\biggl[\frac{1 - \delta_{n,m}}{ |n-m|^\gamma} + \frac{1}{(n+m+1)^\gamma}\biggr]\theta_m
\equiv  
2^{\gamma-1} 
\sum\!\sum\limits\limits_{\hskip-0.5truecm i\neq j} \tilde\theta_i \frac{1}{ |i-j|^\gamma}\tilde\theta_j,
\end{align}
where as before $\frac{1 - \delta_{n,m}}{ |n-m|^\gamma} := 0$ if $n=m$, 
where $i$ and $j$ are odd integers running from $-2N+1$ to $2N-1$, 
and where the truncated sequence $i\mapsto\tilde\theta_i$ is symmetric,
i.e. $\tilde\theta_{-i} = \tilde\theta_i$, with $\tilde\theta_{2n+1} \equiv \theta_n$ for $n\in\{0,...,N-1\}$.
 By the re-arrangement inequality of Hardy--Littlewood--P\'{o}lya \cite{HLP}, see Theorem 3.71 in \cite{HLPbook}, if
the symmetric map $i\mapsto \tilde\theta_i$ is not symmetric decreasing then
{\color{black}
\begin{equation}\label{eq:HLP}
 \sum\limits_i \!\sum\limits_j{}^\sharp{\tilde\theta}_i \frac{1}{ \delta_{i,j} + |i-j|^\gamma}{}^\sharp{\tilde\theta}_j 
> 
 \sum\limits_i \!\sum\limits_j \tilde\theta_i \frac{1}{\delta_{i,j} + |i-j^\prime|^\gamma}\tilde\theta_j,
\end{equation}
where $i\mapsto {}^\sharp{\tilde\theta}_i$ is the symmetric decreasing rearrangement of $i\mapsto\tilde\theta_i$. 
 Note that the diagonal terms (from when $i=j$) contribute $\sum_i  {}^\sharp\tilde\theta_i^2$ to the double sum at 
l.h.s.(\ref{eq:HLP}), and  $\sum_i \tilde\theta_i^2$ to the double sum at r.h.s.(\ref{eq:HLP}).
 Note furthermore that $\ell^2$ norms are invariant under permutations of the elements in the sequence, 
hence under symmetric decreasing rearrangements, so that these diagonal contributions can be purged from (\ref{eq:HLP}).
 Thus, if the symmetric map $i\mapsto \tilde\theta_i$ is not symmetric decreasing then}
\begin{equation}\label{eq:HLPwithoutDIAG}
\sum\!\sum\limits\limits_{\hskip-0.5truecm i\neq j} {}^\sharp\tilde\theta_i \frac{1}{ |i-j|^\gamma}{}^\sharp\tilde\theta_j,
> 
\sum\!\sum\limits\limits_{\hskip-0.5truecm i\neq j} \tilde\theta_i \frac{1}{ |i-j|^\gamma}\tilde\theta_j.
\end{equation}

 Third, since $n\mapsto 2n+1$ is strictly increasing on $\NN_0$, and $N>1$, also the inequality
$\langle {}^\sharp\Theta_N, \mathfrak{D}^{(N)}\, {}^\sharp\Theta_N\big\rangle < 
  \langle \Theta_N, \mathfrak{D}^{(N)}\Theta_N\big\rangle$ holds (under the same hypothesis on $\Theta_N$).
 And since we know that the maximum is positive, we only need to consider $\Theta_N$ for which 
${\big\langle\Theta_N,\widehat{\mathfrak{G}}^{(N)}(\gamma)\,\Theta_N\big\rangle}>0$, so that 
${\big\langle{}^\sharp\Theta_N,\widehat{\mathfrak{G}}^{(N)}(\gamma)\,{}^\sharp\Theta_N\big\rangle}>0$ as well.
 For any non-negative sequence $\Theta_N$ that yields a positive result for the quadratic form (\ref{eq:WIDEhatG}), 
but is locally strictly increasing somewhere, it then follows that 
\begin{equation}\label{eq:rearrangementINEQU}
 \frac{\big\langle\Theta_N,\widehat{\mathfrak{G}}^{(N)}(\gamma)\,\Theta_N\big\rangle}
        {\big\langle\Theta_N,\mathfrak{D}^{(N)}\Theta_N\big\rangle}
 <
 \frac{\big\langle{}^\sharp\Theta_N,\widehat{\mathfrak{G}}^{(N)}(\gamma)\,{}^\sharp\Theta_N\big\rangle}
        {\big\langle{}^\sharp\Theta_N,\mathfrak{D}^{(N)}{}^\sharp\Theta_N\big\rangle}.
\end{equation}

\hspace{-7pt}
The three rearrangement inequalities imply 
that $\Theta^{\mbox{\tiny{\rm opt}}}_N(\gamma) = {}^\sharp\Theta^{\mbox{\tiny{\rm opt}}}_N(\gamma)$.

\hspace{-7pt}
This completes the proof of Proposition~5.\hfill{\bf Q.E.D.}

\noindent
{{\bf Remark~4}: \textsl{The Hardy--Littlewood--P\'{o}lya rearrangement inequality for finite sequences on symmetric intervals
 of integers has been generalized to $\ell^2$ sequences on $\ZZ$ by Pruss \cite{Pruss}. 
 A continuum version for non-negative $L^2(\RR)$ functions is due to F. Riesz \cite{Riesz}; see Theorem 379 in \cite{HLPbook}.}}
\hfill $\square$

\noindent
{{\bf Remark~5}}: \textsl{{
Completely analogously to Proposition 5 it can be shown with the help of Pruss' discretized Riesz inequality that
$\Theta^{\mbox{\tiny{\rm opt}}}(\gamma) = {}^\sharp\Theta^{\mbox{\tiny{\rm opt}}}(\gamma)$.}
 Since in $\ell^2_{\geq0}(\NN_0)$ a constant sequence $\Theta$ has to vanish identically, and thus 
cannot satisfy the normalization of $\Theta$, it follows that the completely positive sequence
$\Theta^{\mbox{\tiny{\rm opt}}}(\gamma)$ is not only also a decreasing sequence, it is strictly decreasing 
infinitely often.}
\hfill $\square$

\noindent
{{\bf Remark~6}:} \textsl{Since $(n,m)\mapsto  \frac{1}{(n+m+1)^\gamma}$ with $(n,m)\in\NN^2_0$ is 
strictly decreasing, separately and jointly so, given any $\Theta\in\ell^2_{\geq0}(\NN_0)$ that is locally strictly increasing
somewhere its decreasing rearrangement ${}^\sharp\Theta$ obeys the  inequality
\begin{equation}
\sum\limits_n \!\sum\limits_m
{}^\sharp\theta_n \frac{1}{(n+m+1)^\gamma}{}^\sharp\theta_m >
\sum\limits_n \!\sum\limits_m \theta_n \frac{1}{(n+m+1)^\gamma}\theta_m.
\end{equation}
{The analogous conclusion does not hold for the form in which $\frac{1}{(n+m+1)^\gamma}$ is replaced by
the Riesz kernel $|n-m|^{-\gamma}$, which is the reason for why
we had to work with the symmetric extension $\widetilde{\Theta}_N$ of $\Theta_N$.}}
\hfill $\square$

 Armed with Proposition~5 we return to the proof of Proposition~4 for when $N>1$ and $\gamma\neq \gamma^\prime$.
\textcolor{black}{We first note that 
since $\Xi^{\mbox{\tiny{\rm opt}}}(\gamma)$ is not the vanishing sequence, the denominator 
of l.h.s.(\ref{eq:expectKgammaWITHgammaNULL}) is $>0$, and thus to prove Proposition 4 for when $N>1$ and $\gamma\neq \gamma^\prime$
it suffices to demonstrate that the numerator of l.h.s.(\ref{eq:rearrangementINEQU}) is $>0$ for \emph{non-vanishing}, non-negative, 
decreasing $\Theta_N\in \RR^N\!;$ for this we may drop the ${}^\prime$ from $\gamma^\prime$.\hspace{-5pt}
}

\textcolor{black}{To demonstrate the positivity of the numerator of l.h.s.(\ref{eq:rearrangementINEQU}) we expand it into a Dirichlet series, viz
\begin{align}\notag
\hspace{-1truecm}
- \sum\limits_{n=0}^{N-1}
 \biggl( {\sum\limits_{k=1}^{n}} \frac{2}{k^\gamma}\!\biggr)  \theta_n^2   
 + \sum\limits_{n=0}^{N-1} \!\sum\limits_{m=0}^{N-1}
\theta_n\biggl[\frac{1 - \delta_{n,m}}{ |n-m|^\gamma} + \frac{1}{(n+m+1)^\gamma}\biggr]\theta_m &
\\  = \label{eq:DIRICHLETexpansionNUM}
\sum\limits_{k=1}^{2N-1}\frac{c_k(\Theta)}{k^\gamma},&
\end{align}
where  the coefficients $c_k(\Theta)$ are given by}
\newpage

\textcolor{black}{
\begin{align}\label{eq:DIRICHLETcoefficients}
c_k(\Theta) &= \ONE_{\{k\leq N-1 \}}\, \sum\limits_{n=k}^{N-1} 2 \big(\theta_{n-k} - \theta_n \big)\theta_n \\
& \notag + \ONE_{\{2\leq k\leq N \}} \, \sum\limits_{n=0}^{k-2} 2 \theta_{k-1-n} \theta_n \\
& \notag +\ONE_{\{ k\,\mbox{\tiny{odd}}\}}\, \theta_{\frac{k-1}{2}}^2,
\end{align}
with $\ONE_{A}$ the indicator function of the set $A$.
 Since the $\theta_n\geq 0$, all contributions from the second and third line at r.h.s.(\ref{eq:DIRICHLETcoefficients}) are manifestly non-negative,
and since the sequence $\Theta_N$ is not only non-negative but also decreasing, it follows that also the contribution from the first
line at r.h.s.(\ref{eq:DIRICHLETcoefficients}) is manifestly non-negative; lastly, since the sequence $\Theta_N$ is not 
only non-negative and decreasing, but also non-vanishing, it follows that $\theta_0^2 >0$, which is contributed by the $k=1$ term in the third line
at r.h.s.(\ref{eq:DIRICHLETcoefficients}). 
 Thus the numerator of l.h.s.(\ref{eq:rearrangementINEQU}) is $>0$ on the set of 
{non-vanishing}, non-negative, decreasing $\Theta_N$. }

The proof of Proposition~4 is complete.\hfill{\bf Q.E.D.}
\smallskip

The proof of Theorem~7  is complete. \hfill{\bf Q.E.D.}
\smallskip

\noindent
\textcolor{black}{
{\bf Remark~7}: \textsl{For our purposes the proof of the positivity of 
l.h.s.(\ref{eq:rearrangementINEQU}) on the set of non-vanishing, non-negative, decreasing sequences suffices.
 An interesting question in its own right is whether there is a positive lower bound to
l.h.s.(\ref{eq:rearrangementINEQU}) on this set.
 We proved that the family of constant non-vanishing sequences is a local minimizer of 
l.h.s.(\ref{eq:rearrangementINEQU}) on this set.
We suspect that these constant sequences in fact are global minimizers on this set, which is also supported by
numerical comparisons of the evaluations of l.h.s.(\ref{eq:rearrangementINEQU}) for a constant sequence with
those for a handful of other pertinent sequences.
 Since the contribution from the first line at r.h.s.(\ref{eq:DIRICHLETcoefficients}) vanishes if 
$\Theta_N$ is a constant sequence, and since those from the second and third line reduce to simple numerical multiples of $\theta_0^2$,
and since $\theta_0^2$ cancels at l.h.s.(\ref{eq:rearrangementINEQU}), 
our suspicion becomes}
}
\smallskip

\noindent
\textcolor{black}{
{\bf Conjecture~3}: \textsl{On the set of non-vanishing, non-negative, decreasing sequences, the l.h.s.(\ref{eq:rearrangementINEQU}) 
is bounded below through}
\begin{equation}
\frac{\big\langle\Theta_N
\widehat{\mathfrak{G}}^{(N)}(\gamma)\,\Theta_N
\big\rangle}
        {\big\langle\Theta_N,\mathfrak{D}^{(N)}\Theta_N\big\rangle}
\geq  
\frac{1}{N^2}\sum_{k=1}^{2N-1} \frac{\min\{k,2N-k\}}{k^\gamma} ,
\end{equation}
with equality holding iff $\Theta_N$ is a constant sequence.
}

\smallskip
\hfill $\square$

\smallskip


\section{A rigorous lower bound on~$\Lambda(P,T)$}\label{sec:lowerLambdaB}

 We here prove Theorem~6.
\smallskip
 
\noindent
\textsl{Proof of Theorem~6}:
 When $\lambda< \Lambda(P,T)$, then  the largest eigenvalue $\kappa$ 
of the linear operator $ \mathfrak{I} -\lambda\mathfrak{K}$ satisfies $\kappa= 1-\lambda\mathfrak{k} >0$.
 In that case the normal state $\underline\Xi\equiv 0$ is linearly stable, i.e.
$Q(\Xi)\geq 0$ on all of $\ell^2(\NN_0)$, with ``$=0$'' if and only if $\Xi=\underline\Xi$.

 Now recall that in the proof of Theorem~$1^+$ we noted that when $\lambda=\Lambda(P,T)$ then
the eigenvalue problem $(\mathfrak{I} -\lambda\mathfrak{K})\Xi_{\mbox{\tiny{\rm opt}}}$ is equivalent to the
fixed point problem  $\mathfrak{C}\big(\tfrac{1}{\Lambda(P,T)}\big)\Xi_{\mbox{\tiny{\rm opt}}}= \Xi_{\mbox{\tiny{\rm opt}}}$,
and the compact operator $\mathfrak{C}\big(\tfrac{1}{\Lambda(P,T)}\big)$ has spectral radius 
$\rho\big(\mathfrak{C}\big(\tfrac{1}{\Lambda(P,T)}\big)\big) =1$.

 The two observations in concert imply that when $\lambda<\Lambda(P,T)$, then 
$\rho\big(\mathfrak{C}\big(\frac1\lambda\big)\big) <1$.
And since $\mathfrak{C}(\eta)$ leaves $\ell^2_{\geq 0}(\NN_0)$ invariant for any $\eta >0$, 
this now means that when $\lambda<\Lambda(P,T)$ then 
$\mathfrak{C}\big(\tfrac1\lambda\big)$ is a contraction mapping on $\ell^2_{\geq 0}$, with $\Xi=0$ as the only fixed point.
 Put differently, if $\lambda<\Lambda(P,T)$, then
\begin{equation}\label{eq:LIPestim}
\big\| \mathfrak{C}\big(\tfrac{1}{\lambda}\big)(\Xi - \Xi')\big\| \leq L[\lambda] \big\|\Xi - \Xi'\big\|,
\end{equation}
with $L[\lambda]:= \rho\big(\mathfrak{C}\big(\tfrac1\lambda\big)\big) <1$ the 
\textsl{Lipschitz constant} for the linear map $\mathfrak{C}\big(\tfrac1\lambda\big):\ell^2_{\geq 0}\to\ell^2_{\geq 0}$. 
\smallskip

\noindent
{\bf Remark~8}: \textsl{The previous paragraph can be rephrased by saying
that linear stability of the normal state $\underline{\Xi}$ against superconducting perturbations
is equivalent to $\mathfrak{C}\big(\tfrac1\lambda\big)$ being a contraction mapping on $\ell^2_{\geq 0}$.}
\hfill $\square$

 We next construct a lower bound $\Lambda^*(P,T)$ on $\Lambda(P,T)$, for any $T > 0$ and $P\in\mathcal{P}$, by showing that 
$\rho\big(\mathfrak{C}\big(\frac{1}{\Lambda^*}\big)\big) < 1$. 
 We accomplish this by arguing as in  \cite{KAYgamma}, invoking two well-known lemmas.
 The first one is

\smallskip
\noindent
{\bf Lemma~1}: \textsl{Let $T> 0$ and $P$ be given.  
 Then 
\begin{equation}
\rho(\mathfrak{C}\big(\eta\big) )\leq 
\rho\Big(\big(\eta \mathfrak{I} +\mathfrak{K}_1\big)^{-1}\Big)
\rho\big( \mathfrak{K}_2 + \mathfrak{K}_3\big).
\end{equation}
}

 Recall that $\big(\eta\mathfrak{I} +\mathfrak{K}_1\big)^{-1}$ is a diagonal operator with all positive entries, 
and that $\mathfrak{K}_1$ has non-negative entries including 0. 
 Hence, 
\begin{equation}\label{eq:diagEST}
\hspace{-1truecm}
\rho\Big(\!\big(\tfrac1\lambda \mathfrak{I} +\mathfrak{K}_1\big)^{-1}\Big) = 
\max_{n\geq 0} 
\Big[\frac1\lambda + \frac{2}{2 n + 1} {\textstyle\sum\limits_{k=1}^{n}} \int_0^\infty\!\!\frac{\varpi^2}{\varpi^2 + k^2} P(d\omega)\Big]^{-1}
\!\!=\lambda.
\end{equation}
 Since the operator $ \mathfrak{K}_2 + \mathfrak{K}_3$ is independent of $\lambda$, we now arrive at the following conclusion.

\smallskip
\noindent 
{\bf Proposition~6}: \textsl{Let $T> 0$ and $P\in\mathcal{P}$ be given. Suppose 
\begin{equation}\label{eq:PROPhypo}
\lambda \leq   \tfrac{1}{\rho\big( \mathfrak{K}_2 + \mathfrak{K}_3\big)}.
\end{equation}
 Then  $\lambda \leq \Lambda(P,T)$.}
\smallskip

\noindent
\textsl{Proof}: By Lemma~1 and by (\ref{eq:diagEST}), and by the hypothesis (\ref{eq:PROPhypo}) of the Proposition, we have that
\begin{equation}\label{eq:specRADchainEST}
\rho(\mathfrak{C}\big(\tfrac1\lambda\big) ) \leq \lambda
\rho\big( \mathfrak{K}_2 + \mathfrak{K}_3\big) 
\leq 1 = \rho\big(\mathfrak{C}\big(\tfrac{1}{\Lambda(P,T)}\big) \big) .
\end{equation}
Since $\eta\mapsto \rho(\mathfrak{C}\big(\eta\big))$ is strictly monotonically decreasing, the claim of the Proposition follows.
\hfill {\textbf{Q.E.D.}}

\noindent
{\bf Corollary~5}:
 \textsl{Under the hypotheses of Proposition~6, we have}
\begin{equation}\label{eq:TAUcBOUNDup}
\Lambda(P,T)
 \geq  \tfrac{1}{\rho\big( \mathfrak{K}_2 + \mathfrak{K}_3\big)}.
\end{equation}
\smallskip

The second well-known fact is
\smallskip

\noindent
{\bf Lemma~2}: \textsl{Let $T> 0$ and $P\in\mathcal{P}$ be given.  
 Then 
\begin{equation}\label{eq:rho1rho2}
\rho\big( \mathfrak{K}_2 + \mathfrak{K}_3\big)
\leq 
\rho\big( \mathfrak{K}_2\big) + \rho\big(\mathfrak{K}_3\big).
\end{equation}
}
\smallskip

\noindent
{\bf Corollary~6}:
 \textsl{Under the hypotheses of Proposition~6, we have}
\begin{equation}\label{eq:TAUcBOUNDupUP}
\Lambda(\varpi)
 \geq  \tfrac{1}{\rho\big( \mathfrak{K}_2\big) + \rho\big(\mathfrak{K}_3\big)}.
\end{equation}
\smallskip

We now estimate the two spectral radii at r.h.s.(\ref{eq:rho1rho2}) from above in terms
of the spectral radii of $\mathfrak{G}_2(2)$ and $\mathfrak{G}_3(2)$ that we estimated in \cite{KAYgamma}.

\smallskip
\noindent
{\bf Proposition~7}: \textsl{Let $T> 0$ and $P\in\mathcal{P}$ be given.  Then,  with $\varpi=\omega/2\pi T$, for $j\in\{2,3\}$ we have} 
\begin{equation}\label{eq:PROPnew}
\rho\big( \mathfrak{K}_j(P,T)\big) \leq \rho\big( \mathfrak{G}_j(2)\big) \int_0^\infty \varpi^2 P(d\omega).
\end{equation}
\smallskip

\noindent
\textsl{Proof}: All kernels of the operators 
$\mathfrak{K}_j(P,T)$, $j\in\{2,3\}$, are of the type $\int_0^\infty\!\frac{\varpi^2}{\varpi^2+k^2}P(d\omega)>0$
for $k\in\NN$, while all kernels of $ \mathfrak{G}_j(\gamma)$, $j\in\{2,3\}$, are of the type $\frac{1}{k^\gamma}>0$
for $k\in\NN$.
 So the spectral radii of all these operators are achieved on the positive cone $\ell^2_{\geq0}(\NN_0)$.
 Since $\varpi^2+k^2 > k^2$, the claim of the proposition follows. \hfill{\bf Q.E.D.}

\bigskip
Now we recall that $\rho\big( \mathfrak{G}_j(2)\big)$ for $j\in\{2,3\}$ has been estimated from above in \cite{KAYgamma},
see their Propositions~5 and~6. 
 These estimates prove the weaker upper estimate of $T_c(P,T)$ registered after the statement of Theorem~6, 
where the term $\mathfrak{k}^{(1)}(P,T) = \int_0^\infty
\frac{\varpi^2}{1+\varpi^2}P(d\omega)$ in  (\ref{eq:kSTAR}) 
is replaced by the weaker $\int_0^\infty\varpi^2P(d\omega)$.
 This almost proves our Theorem~6.

 Finally, closer inspection of the proof of Proposition~6 in \cite{KAYgamma} reveals that the estimate $\varpi^2+1 > 1$, 
used in the proof of Proposition~7 above, is not needed to arrive at an analog of inequality (100) in \cite{KAYgamma}, 
and avoiding the unnecessary estimate $\varpi^2+1 > 1$ for this particular contribution to the upper bound on 
$\rho\big( \mathfrak{K}_3\big)$ gives the bound stated in Theorem~6.  
\hfill{\bf Q.E.D.}

\section{From $\Lambda(P,T)$ to $T_c^{}(\lambda,P)$}\label{sec:TcA}

We here prove Proposition~1, and then Theorem~2 and Corollary~1. 
\smallskip

\noindent
\textsl{Proof of Proposition~1}: 
By its definition, $\mathfrak{k}^{(N)}(P,T)$ is given by the variational principle
\begin{equation}\label{eq:hVPtrialN}
\mathfrak{k}^{(N)}(P,T) : = 
\max_{\Xi_N} \frac{\big\langle\Xi_N,\mathfrak{K}^{(N)}(P,T)\,\Xi_N\big\rangle}
        {\big\langle\Xi_N,\Xi_N\big\rangle}.
\end{equation}
 With the normalization ${\big\langle\Xi_N,\Xi_N\big\rangle}=1$, and with the optimizer denoted 
$\Xi^{\mbox{\tiny{\rm opt}}}_N(P,T)$, we thus have
\begin{equation}\label{eq:kNagain}
\mathfrak{k}^{(N)}(P,T)  = 
\big\langle\Xi^{\mbox{\tiny{\rm opt}}}_N(P,T),\mathfrak{K}^{(N)}(P,T)\,\Xi^{\mbox{\tiny{\rm opt}}}_N(P,T)\big\rangle.
\end{equation}
 Since all kernels of $\mathfrak{K}^{(N)}(P,T)$ depend on $T$ only through analytic functions of $T^2$, 
by first order perturbation theory \cite{Kato} we have
\begin{equation}\label{eq:kNprime}
\tfrac{\partial}{\partial T^2}
\mathfrak{k}^{(N)}(P,T)  = 
\big\langle\Xi^{\mbox{\tiny{\rm opt}}}_N(P,T),\big[\tfrac{\partial}{\partial T^2}\mathfrak{K}^{(N)}(P,T)\big]\,\Xi^{\mbox{\tiny{\rm opt}}}_N(P,T)\big\rangle.
\end{equation}
 We want to show that r.h.s.(\ref{eq:kNprime})$<0$.
 So far we can show this only if $T \geq T_*(P)$ with $T_*(P) \leq \frac{\overline\Omega(P)}{2\sqrt{2}\pi}$.
 To do so we recycle the strategy of proving Proposition~4.

 As in the proof of Proposition~4, we need the analog of Proposition~5, now for the standard Eliashberg model.
\smallskip

\noindent
{\bf Proposition~8}: \textsl{For all $N\in\NN_0$, the map} 
$n\mapsto \frac{1}{\sqrt{2n+1}}\big(\Xi^{\mbox{\tiny{\rm opt}}}_N(P,T)\big)_n$ 
\textsl{is positive and decreasing.}
\smallskip

\noindent
\textsl{Proof}: 
 The proof of Proposition~8 is essentially verbatim to the proof of Proposition~5 and does not need to be 
stated in detail. 
 It should suffice to recall that $\mathfrak{K}^{(N)}= - \mathfrak{K}^{(N)}_1 + \mathfrak{K}^{(N)}_2 + \mathfrak{K}^{(N)}_3$
in complete analogy to $\mathfrak{G}^{(N)}= - \mathfrak{G}^{(N)}_1 + \mathfrak{G}^{(N)}_2 + \mathfrak{G}^{(N)}_3$, 
with kernels of the type $\int_0^\infty\!\!\frac{\varpi^2}{\varpi^2+j^2}P(d\omega)$ 
taking over the role of the kernels of the type $\frac{1}{j^\gamma}$.
 Note that both types of kernels are positive, and they decrease when the index $j\in\NN$ is increased. 

This suffices to prove Proposition~8. \hfill {\textbf{Q.E.D.}}
\smallskip

 Next we register the componentwise actions of $\tfrac{\partial}{\partial T^2}\mathfrak{K}^{(N)}_j(P,T)$ as $N$-frequency 
truncations of the componentwise actions of $\tfrac{\partial}{\partial T^2}\mathfrak{K}_j(P,T)$, given by
the $P$-averages of $-{\omega^2}/{4\pi^2 T^4}$ times the $N$-frequency truncations of the componentwise actions 
 \begin{align}
\label{eq:HopsCompsONEprime}
\hspace{-1.1truecm}
\biggl(\!\frac{d}{d\varpi^2}\mathfrak{H}_1(\varpi)\Xi\!\biggr)_{\!n}\! = & 
 \biggl[\frac{2}{2 n + 1} {\sum\limits_{k=1}^{n}} \frac{k^2}{(\varpi^2 + k^2)^2}\biggr] \xi_n \,, \\
\label{eq:HopsCompsTWOprime}
\hspace{-1.1truecm}
\biggl(\!\frac{d}{d\varpi^2}\mathfrak{H}_2(\varpi)\Xi\!\biggr)_{\!n}\! = & 
 \sum\limits_{m} 
\biggl[\frac{1}{\sqrt{2 n + 1}}\,\frac{(n-m)^2}{(\varpi^2+(n-m)^2)^2}\,\frac{1}{\sqrt{2 m+1}}\biggr]\xi_m,\!\!\!\!\\
\label{eq:HopsCompsTHREEprime}
\hspace{-1.1truecm}
\biggl(\!\frac{d}{d\varpi^2}\mathfrak{H}_3(\varpi)\Xi\!\biggr)_{\!n}\! = & 
 \sum\limits_m \biggl[\frac{1}{\sqrt{2 n + 1}}\, \frac{(n+m+1)^2}{(\varpi^2 +(n+m+1)^2)^2}\, \frac{1}{\sqrt{2 m + 1}}\biggr]\xi_m\,.
\end{align}

\noindent
{\bf Lemma~3}: \textsl{Let $\varpi>0$.
 Then the map $j\mapsto G_\varpi(j):=\frac{j^2}{(\varpi^2+j^2)^2}$, $j\in\NN$, is decreasing iff 
$\varpi\leq \sqrt{2}$.}
\smallskip

\noindent
\textsl{Proof}: Elementary calculus reveals that the rational function $x\mapsto \frac{x^2}{(\varpi^2+x^2)^2}$ on $\RR_+$
increases strictly monotonically for $x\in[0,\varpi]$, reaches a unique maximum at $x=\varpi$, and decreases strictly
monotonically to 0 for $x\in[\varpi,\infty)$.
 This already implies that $G_\varpi(j)$, $j\in\NN$, decreases with increasing $j$ if $\varpi\leq 1$.
 Yet, since $j\in\NN$, we can increase $\varpi$ a little bit beyond 1 so long as $G_\varpi(2)\leq G_\varpi(1)$.
 This is only feasible when the maximum of $G_\varpi(x), x\in\RR_+$, is located at some $x$ below $\sqrt{2}$.
 The limiting situation occurs for $\varpi=\sqrt{2}$, with $G_{\sqrt{2}}(2)=G_{\sqrt{2}}(1)$.
\hfill{\bf Q.E.D.}
\smallskip

We note that when $\varpi=\sqrt{2}$, so that $G_{\sqrt{2}}(2)=G_{\sqrt{2}}(1)$,
one still has $G_{\sqrt{2}}(j+1)< G_{\sqrt{2}}(j)$ for all $j>1$. 
 \smallskip

\noindent
{\bf Corollary~7}:
 \textsl{Let $T\geq {\overline\Omega(P)}/{2\sqrt{2}\pi}$.  
Then r.h.s.(\ref{eq:kNprime})$<0$.}
\smallskip

\noindent
\textsl{Proof}: Since by general hypothesis on $P$ its support is contained in $[0,\overline\Omega(P)]$, 
the hypothesis of Corollary~7 that $T\geq {\overline\Omega}/{2\sqrt{2}\pi}$ implies that ${\omega}/{2\pi T} = \varpi\leq \sqrt{2}$
for all $\omega \leq \overline\Omega(P)$.

 Thus, having Proposition~8 and Lemma~3, we can follow the strategy of the
proof of Proposition~4, recall the normalization 
${\big\langle\Xi^{\mbox{\tiny{\rm opt}}}_N,\Xi^{\mbox{\tiny{\rm opt}}}_N\big\rangle}=1$ with 
$\Xi^{\mbox{\tiny{\rm opt}}}_N\equiv \Xi^{\mbox{\tiny{\rm opt}}}_N(P,T)$, and conclude that for $\varpi\in[0,\sqrt{2}]$ we have
\begin{align}\label{eq:HprimeNboundLOW}
 \big\langle\Xi^{\mbox{\tiny{\rm opt}}}_N,
\big[\tfrac{d}{d\varpi^2}\mathfrak{H}^{(N)}(\varpi)\big]\,\Xi^{\mbox{\tiny{\rm opt}}}_N\big\rangle
\color{black}{>0}.
\end{align}
 Now multiplying inequality (\ref{eq:HprimeNboundLOW}) by $-\omega^2/4\pi^2T^4$,
then averaging over $P(d\omega)$, reverses the inequality in (\ref{eq:HprimeNboundLOW}).
 
This proves Corollary~7. \hfill {\textbf{Q.E.D.}}

The proof of Proposition~1 is complete. \hfill {\textbf{Q.E.D.}}
\smallskip

\noindent
{\color{black}
{\bf Remark~9}: \textsl{Following up on Remark 8, we suspect that 
\begin{align}\label{eq:HprimeNboundLOWbetter}
 \big\langle\Xi^{\mbox{\tiny{\rm opt}}}_N,
\big[\tfrac{d}{d\varpi^2}\mathfrak{H}^{(N)}(\varpi)\big]\,\Xi^{\mbox{\tiny{\rm opt}}}_N\big\rangle
\geq 
{\big\langle\overline{\Theta}_N,
\big[\tfrac{d}{d\varpi^2}\widehat{\mathfrak{H}}^{(N)}(\varpi)\big]\,\overline{\Theta}_N\big\rangle,}
\end{align}
}where $\widehat{\mathfrak{H}}^{(N)}(\varpi)$ is obtained from ${\mathfrak{H}}^{(N)}(\varpi)$ 
analogously to how $\widehat{\mathfrak{G}}^{(N)}(\gamma)$ is obtained from ${\mathfrak{G}}^{(N)}(\gamma)$
in the proof of Proposition~4, and
where $\overline\Theta_N$ is the  constant vector $(\frac1N,...,\frac1N)$; recall that 
${\big\langle\overline\Theta_N,\mathfrak{D}^{(N)}\overline\Theta_N\big\rangle} =  1$.
 Note that equality in (\ref{eq:HprimeNboundLOWbetter}) holds iff $N=1$.
 R.h.s.(\ref{eq:HprimeNboundLOWbetter}) is a weighted sum of the kernels  $\frac{k^2}{(k^2+\varpi^2)^2}$ with
$k\in\{1,...,2N-1\}$, yielding}
\begin{align}\label{eq:HprimeNboundLOWrewriteBETTER}
& \Big\langle\overline{\Theta}_N,
\Big[\frac{d}{d\varpi^2}\widehat{\mathfrak{H}}^{(N)}(\varpi)\Big]\,\overline{\Theta}_N\Big\rangle
=  \\ \label{eq:HprimeNboundLOWrewriteISpos}
& \qquad \frac{1}{N^2}{\textstyle\sum\limits_{k=1}^{2N-1}}\frac{k^2}{(k^2+\varpi^2)^2} \min\{k,2N-k\}
> 0.
\end{align}
\hfill $\square$
\smallskip

\noindent
\textsl{Proof of Theorem~2}:
 The strict monotonicity of $T\mapsto\Lambda^{(N)}(P,T)$, given $P$, for  $T\geq {\overline\Omega(P)}/{2\sqrt{2}\pi}$ and 
all $N\in\NN$, implies via the convergence $\Lambda^{(N)}(P,T)\to \Lambda(P,T)$ as $N\to\infty$, also
the monotonicity of $T\mapsto\Lambda(P,T)$ if $T\geq {\overline\Omega(P)}/{2\sqrt{2}\pi}$.
 Since $T\mapsto\Lambda(P,T)$ is analytic for each $P\in\mathcal{P}$, it cannot have a constant piece, for this would violate
its lower bound $\Lambda^*(P,T)$; hence $T\mapsto\Lambda(P,T)$ is strictly monotonic increasing, too.
 Yet ${\overline\Omega(P)}/{2\sqrt{2}\pi}$ is merely the lower limit of $T$ values for which our reasoning 
establishes the negativity of $\tfrac{\partial}{\partial T^2}\mathfrak{k}(P,T)$.
 Our argument does not reveal ``how negative'' $\tfrac{\partial}{\partial T^2}\mathfrak{k}(P,T_*(P))$ is.
 Yet it is $<0$, and so by continuity 
$\tfrac{\partial }{\partial T^2}\mathfrak{k}(P,T))<0$ also in some left neighborhood of ${\overline\Omega(P)}/{2\sqrt{2}\pi}$.

The proof of Theorem~2 is complete. \hfill {\textbf{Q.E.D.}}
\newpage

\noindent
\textsl{Proof of Corollary~1}: Having proved Proposition~1 and also Theorem~4, the explicit bound (\ref{eq:lambdaSUBstarFOUR}) 
on the quantity $\lambda_*(P)$ defined in Theorem~2 follows right away.
 \hfill{\bf Q.E.D.}
\smallskip

 We close this section with some comments on practical matters.
 While the bound (\ref{eq:lambdaSUBstarFOUR}) is explicit, its evaluation still requires the $P$-averages of 
$\frac{\omega^2}{\omega^2 + 4\pi^2T^2 j^2}$ for $j\in\{1,...,7\}$, which in general will require numerical quadratures for each 
$T$ value of interest. 
 A weaker but more user-friendly estimate on $\lambda_*(P)$ is 
\begin{equation}\label{eq:lambdaSUBstarONE}
\lambda_*(P) <\frac{1}{\mathfrak{k}^{(1)}\big(P,T_*(P)\big)}.
\end{equation}
 Since $P$ is compactly supported in $[0,\overline\Omega(P)]$, we furthermore have the estimate
\begin{equation}
\mathfrak{k}^{(1)}(P,T) \geq \tfrac{1}{{\overline\Omega(P)}^2 +(2\pi T)^2}\int_0^\infty\!\!\omega^2 P(d\omega).
\end{equation}
 Now substituting $T_*(P)$ for $T$ yields the upper estimate
\begin{equation}\label{eq:LambdaSUBstarEASY}
\lambda_*(P) \leq \frac32 \frac{{\overline\Omega(P)}^2}{\big\langle\omega^2\big\rangle},
\end{equation}
which requires only knowledge of $\overline\Omega(P)$ and the computation of $\big\langle\omega^2\big\rangle$.

\section{Upper and lower bounds on~$T_c(\lambda,\Omega)$}\label{sec:TcB}

 The validity of Corollary~4 is obvious, so nothing needs to be added here about these explicit but relatively weak bounds
on $T_c(\lambda,P)$.

 Our better bounds on $T_c(\lambda,P)$ are not explicitly given as functions of $\lambda$ and $P$.
 However, since their inverse functions $\lambda = \Lambda^{(N)}(P,T)$, $N\in\{1,2,3,4\}$, respectively
$\lambda = \Lambda^*(P,T)$, are explicitly given in section~\ref{sec:mainRESULTS}, for any $P$, 
one can conveniently visualize the bounds on $T_c(\lambda,P)$ that 
we proved to exist for $T > T_*(P)$ (and which reasonably are expected to exist for all $T>0$).
 All that needs to be done is to graph the $\Lambda^{(N)}(P,T)$ and $\Lambda^*(P,T)$ in a $(\lambda,T)$ diagram, given
a desirable choice of $P$.
 For each $P$ this still requires one or more $T$-dependent quadratures to be carried out, in most cases by numerical 
algorithm unless by good fortune those $P$ integrals can be carried out in terms of known functions. 
 One such fortunate case is the non-dispersive limit, with $P(d\omega) = \delta(\omega-\Omega)d\omega$, for fixed $\Omega>0$;
this case merits a discussion of its own and will be featured in paper III of our series. 

 In the remainder of this section we confine ourselves to adding a mathematical result for general $P$ that we announced in 
section \ref{sec:mainRESULTS}.

\subsection{The lower bound $\textstyle{T_c^{(2)}(\lambda,P)}$}\label{sec:TcTWO}

 As announced in section  \ref{sec:mainRESULTS}, 
we here supply the proof of  monotonicity for the map $T\mapsto \Lambda^{(2)}(P,T)$, given $P$,
which implies its invertibility and the monotonicity of its inverse $\lambda\mapsto T_c^{(2)}(\lambda,P)$, a better lower
bound on $T_c(\lambda,P)$ than $T^\flat(\lambda,P)$.
 While it does not seem to have a closed form expression in known functions,
the lower bound $T_c^{(2)}(\lambda,P)$ is defined for all $\lambda\geq \lambda_2^{}$, as we will see now.

 For the $2\times2$ matrix given by the $P$-average of the upper leftmost $2\times2$ block of r.h.s.(\ref{eq:Hfour}),
the largest eigenvalue $\mathfrak{k}^{(2)}(P,T)$ is given by (\ref{eq:kTWO}), with
\begin{equation}\label{eq:traceKtwo}
{\rm tr}\,\mathfrak{K}^{(2)} = \frac13 \big\langle{[\![} 1{]\!]} + {[\![} 3{]\!]}\big\rangle
\end{equation}
and
\begin{equation}\label{eq:determinantKtwo}
\det\mathfrak{K}^{(2)} = - \frac13 
\left(\!\big\langle{[\![} 1{]\!]} + {[\![} 2{]\!]}\big\rangle^2\! 
+ \big\langle{[\![} 1{]\!]}\big\rangle
 \!\left\langle 2  {[\![} 1{]\!]} - {[\![} 3{]\!]}\right\rangle\! \right),
\end{equation}
where ${[\![}n{]\!]}(\varpi):= \frac{\varpi^2}{n^2 +\varpi^2}$ for $n\in\NN$, with $\varpi=\omega/2\pi T$, 
and where the angular brackets indicate $P(d\omega)$-averages.

 Note that (\ref{eq:traceKtwo}) reveals that ${\rm tr}\,\mathfrak{K}^{(2)}>0$;
note furthermore that $n\mapsto \frac{\varpi^2}{n^2 +\varpi^2}>0$ is strictly decreasing with increasing $n\in\NN$,
and so (\ref{eq:determinantKtwo}) reveals that $\det\mathfrak{K}^{(2)}<0$. 
 This shows that (\ref{eq:kTWO}) is well-defined; incidentally, it also vindicates the choice of the $+$ sign in front of the $\surd$ term
in (\ref{eq:kTWO}), for choosing a $-$ sign instead would not produce a positive eigenvalue.

 Taking the reciprocal of (\ref{eq:kTWO}) yields the upper bound $\Lambda^{(2)}(P,T)$ on $\Lambda(P,T)$, explicitly
\begin{align}\label{eq:LambdaTWOexpl}
 & \hspace{-1truecm} \Lambda^{(2)} =  \\   \notag & \hspace{-0.8truecm}
\frac{6}
     {\big\langle{[\![} 1{]\!]} + {[\![} 3{]\!]}\big\rangle  +
 \sqrt{\big\langle{[\![} 1{]\!]} + {[\![} 3{]\!]}\big\rangle^2\! + 
 12\left(\! \big\langle{[\![} 1{]\!]} + {[\![} 2{]\!]}\big\rangle^2\!
+\big\langle {[\![} 1{]\!]}\big\rangle \big\langle 2 {[\![} 1{]\!]} - {[\![} 3{]\!]}\big\rangle\! \right)}}\,.
\end{align}
 The map $(P,T)\mapsto \Lambda^{(2)}(P,T)$ is readily seen to be jointly continuous.
 We now show that it is also strictly increasing when $T>0$ increases from 0 to $\infty$, given any $P\in\mathcal{P}$.

 Indeed, with $\varpi=\omega/2\pi T$ it is manifest that for any $n\in\NN$, 
the map $T \mapsto {[\![} n{]\!]}(\varpi)\equiv  \frac{\omega^2}{4n^2\pi^2T^2 +\omega^2}$ 
is continuous and monotonically decreasing from 1 to 0 as $T$ runs from $0$ to $\infty$, for each $\omega>0$.
 Thus $\langle{[\![} 1{]\!]} + {[\![} 2{]\!]}\rangle$ and
$\langle{[\![} 1{]\!]} + {[\![} 3{]\!]}\rangle$ and their squares
are bounded, continuous, and monotonically decreasing as $T$ runs from $0$ to $\infty$.
 The only term that could cause problems is the $- \langle{[\![} 3{]\!]}\rangle$ contribution at r.h.s.(\ref{eq:LambdaTWOexpl}),
which increases with $T$.

 After investigating first the $T$ dependence of $\langle{[\![} 1{]\!]}\rangle\!\left\langle2  {[\![} 1{]\!]} - {[\![} 3{]\!]} \right\rangle$ 
and finding that it is not monotonic, taking into account also the additive term 
$\big\langle{[\![} 1{]\!]} + {[\![} 2{]\!]}\big\rangle^2$ yields success. 
 Indeed, rewriting 
$\big\langle{[\![} 1{]\!]} + {[\![} 2{]\!]}\big\rangle^2\!
+\big\langle {[\![} 1{]\!]}\big\rangle \big\langle 2 {[\![} 1{]\!]} - {[\![} 3{]\!]}\big\rangle
= \big\langle {[\![} 2{]\!]}\big\rangle^2 + 
 \big\langle {[\![} 1{]\!]}\big\rangle 
\Big( 3  \big\langle {[\![} 1{]\!]}\big\rangle + 2  \big\langle {[\![} 2{]\!]}\big\rangle -  \big\langle {[\![} 3{]\!]}\big\rangle \Big)$
and noting that $ \big\langle {[\![} 2{]\!]}\big\rangle^2$ and $ \big\langle {[\![} 1{]\!]}\big\rangle$ are monotonically strictly
decreasing with $T$, given $P$, and that the term between big round parentheses that multiplies
$\big\langle {[\![} 1{]\!]}\big\rangle$ is positive (by the decrease of $n^2\mapsto \frac{\varpi^2}{n^2+\varpi^2}$),
it remains to check the $T$ dependence of the term between big round parentheses that multiplies $\big\langle {[\![} 1{]\!]}\big\rangle$.
 With the help of Maple one finds 
(with $C_n\in\NN$, $n\in\{1,...,5\}$)\footnote{For the sake of completeness, we state the numerical values of the $C_n$ explicitly 
here: $C_1=4392$, $C_2=3888$, $C_3=1370$, $C_4=148$, and $C_5=2$.} 
that
\begin{align}\label{eq:MAPLEderiv}
\tfrac{\partial}{\partial T^2}
\Big(3\big\langle {[\![} 1{]\!]}\big\rangle + 2\big\langle {[\![} 2{]\!]}\big\rangle -  \big\langle {[\![} 3{]\!]}\big\rangle \Big)
(P,T)
&= \\ \label{eq:MAPLEderivESTIM}
- \int_0^\infty \frac{\sum_{n=1}^5 C_n \omega^{2n}(4\pi^2 T^2)^{5-n}}{\prod_{j=1}^3 \left(4j^2\pi^2T^2 +\omega^2\right)^2} P(d\omega)
&<0,
\end{align}
which was to be shown.

\smallskip
 We summarize our result in
\smallskip

\noindent
{\bf Proposition~9}
\textsl{Given any $P\in\mathcal{P}$, 
the map $T\mapsto\lambda=\Lambda^{(2)}(P,T)$ is invertible, as announced in section~\ref{sec:mainRESULTS}.
 The inverse function is the lower critical-temperature bound $T_c^{(2)}(\lambda,P)$ that 
continuously and monotonically increases in $\lambda$ on its domain of definition 
$\lambda\in[\lambda_2,\infty)$, with $\lambda_2$ given by (\ref{eq:lambdaN}) for $N=2$; viz.}
\begin{align}\label{eq:LambdaTWOexplEVALinfty}
 \lambda_2 =  \tfrac{3}{5}= 0.6\,. 
\end{align}

 We remark that even in the simplest choice, $P(d\omega)=\delta(\omega-\Omega)d\omega$,
the inversion of $T\mapsto\lambda=\Lambda^{(2)}(P,T)$ is not expressible in closed algebraic form.

\section{Summary and Outlook}\label{sec:sumANDout}

\subsection{Summary}

 In this paper we rigorously studied the phase transition between normal and superconductivity
in the standard Eliashberg theory in which the effective electron-electron interactions are mediated 
by generally dispersive phonons, with {Eliashberg} spectral density function $\alpha^2\!F(\omega)$ that is $\propto\omega^2$
for small $\omega$ and vanishes for sufficiently large $\omega>\overline\Omega$; it also defines the standard
electron-phonon coupling strength $\lambda := 2\int_{\RR_+}\!\!\alpha^2\!F(\omega)\frac{d\omega}{\omega}$.

 After a suitable rescaling with $\lambda$ the standard Eliashberg model is asymptotic to the $\gamma$ model at $\gamma=2$
when $\lambda\to\infty$.
 The $\gamma$ model was studied in our previous paper \cite{KAYgamma}, and the results obtained there were
convenient ingredients to prove several results of the present paper, too.
 Several other results that we proved in the present paper are based on entirely new arguments, though.

 Defining a formal probability measure $P$ through $2\alpha^2\!F(\omega)\frac{d\omega}{\omega}=:\lambda P(d\omega)$,
we in this paper proved that the normal and the superconducting regions in
the positive $(\lambda,P,T)$ cone are both simply connected, and separated 
by a critical surface $\mathscr{S}_c$ that is a graph over the positive $(P,T)$ cone, given by
a function $\lambda = \Lambda(P,T)$.
 This is the content of our Theorem~1.
 We furthermore proved that $\Lambda(P,T) = 1/\mathfrak{k}(P,T)$, where $\mathfrak{k}(P,T)>0$ is the largest
eigenvalue of an explicitly constructed compact operator $\mathfrak{K}(P,T)$ on $\ell^2(\NN_0)$, where $\NN_0$ is 
the set of non-negative integers that enumerates the positive Matsubara frequencies.
 This is stated in Theorem~2.

 Since a compact operator on a separable Hilbert space can be arbitrarily closely approximated by truncating it to finite-dimensional
subspaces, in this case spanned by the first $N$ positive Matsubara frequencies, we obtained 
from our variational principle a strictly monotonically decreasing sequence of rigorous upper bounds on $\Lambda(P,T)$,
the first four of which we have computed explicitly in closed form, though still involving quadratures of 
$\frac{\omega^2}{\omega^2+ (2n\pi T)^2}$ w.r.t. $P(d\omega)$ for $n\in\{1,...,2N-1\}$.
 This is the content of our Theorem~4.

 In Theorem~5 we explicitly stated the bounds $\Lambda^{(N)}(P,T)$ in the limit $T\to 0$.
 Interestingly these limits are independent of the choice of $P$.

 Through spectral estimates of $\mathfrak{k}(P,T)$ from above we also rigorously obtained an explicit 
lower bound on $\Lambda(P,T)$, involving only the quadratures of $\omega^2$ and of
$\frac{\omega^2}{\omega^2+ (2\pi T)^2}$ w.r.t. $P(d\omega)$; see our Theorem~6.
 At the expense of less accuracy we also obtained weaker upper and lower bounds on $\Lambda(P,T)$ involving only
the quadrature of $\omega^2$ w.r.t. $P(d\omega)$, and the maximum frequency $\overline\Omega$, as a corollary to Theorem~6.
 
 Physical intuition, based on empirical evidence, suggests that the phase transition can be characterized in terms of a 
\textsl{critical temperature} $T_c(\lambda,P)$, which is equivalent to saying the critical surface $\mathscr{S}_c$ 
is a graph over the positive $(\lambda,P)$ cone.
 This in turn is equivalent to saying that, given $P$, the map $T\mapsto\Lambda(P,T)$ is strictly monotone, hence invertible to yield
$T=  \Lambda^{-1}(\lambda,P)\equiv T_c(\lambda,P)$.
 In this paper we were able to prove that all our upper approximations to 
the map $T\mapsto\Lambda(P,T)$, and that map itself, 
are strictly monotone decreasing for $T\in[T_*(P),\infty]$, with $T_*(P) \leq \frac{\overline\Omega(P)}{2\sqrt{2}\pi}$.
 This is the content of Theorem~2.

 The restriction of the monotonicity result to $T\in[T_*(P),\infty]$ is due to the limitations of our technique of proof and
presumably not intrinsic to the model.
 In this vein, we do expect that the map $T\mapsto\Lambda(P,T)$ is strictly monotone increasing 
for all $T\in[0,\infty)$, given any $P\in\mathcal{P}$.
 To prove this is a worthy goal for future works.

 Since the explicit fourth upper bound on $\Lambda(P,T)$ yields $\lambda_*(P) <\frac{1}{\mathfrak{k}^{(4)}\big(P,T_*(P)\big)}$,
what we just wrote proves that a unique critical temperature $T_c(\lambda,P)$ in the standard Eliashberg model is 
mathematically well-defined in terms of the untruncated linearized Eliashberg gap equations whenever $\lambda>\lambda_*(P)$.
 A weaker but rather explicit estimate is obtained with the weaker first upper bound on $\Lambda(P,T)$, further estimated
itself in terms of the expected value of $\omega^2$ w.r.t. $P(d\omega)$ and in terms of $\overline\Omega$.
 This has yielded the estimate 
$\lambda_*(P) \leq \frac32 \frac{{\overline\Omega(P)}^2}{\langle\omega^2\rangle}$.

 Also $\lambda_*$ is not a sharp boundary but a consequence of our method of proof.
 While mathematically desirable to prove the existence of a unique $T_c(\lambda,P)$ in the Eliashberg model for all
$\lambda>0$, from a theoretical physics perspective the range 
$\lambda> \frac32 \frac{{\overline\Omega(P)}^2}{\langle\omega^2\rangle}$
would seem to cover all cases of interest so far.

 Furthermore we proved that asymptotically for large $\lambda$ one has $T_c(\lambda,P)\sim C\sqrt{\langle\omega^2\rangle\lambda}$, 
with $C=\frac{1}{2\pi}\sqrt{\mathfrak{g}(2)}=0.1827262477...$, where $\mathfrak{g}(2)$ is the spectral radius of a 
compact operator $\mathfrak{G}(2)$ associated with the $\gamma$ model for $\gamma=2$.
 This is expressed in Theorem~7 and its corollary.

 With the existence of a unique critical temperature secured for most situations of interest, we have the following 
interesting application of our results.
\smallskip

\begin{quote}
 It is known that if one measures $\alpha^2\!F(\omega)$ reasonably accurately in a laboratory experiment, then $\lambda$ 
is obtained by a single numerical quadrature via (\ref{eq:lambda}).
 Yet suppose that only $P(d\omega)$ can reasonably accurately be measured for certain materials featuring a superconductivity transition 
temperature $T_c$, and suppose that also $T_c$ has been measured in the laboratory.
 Then numerical quadrature of $\omega^2$ and of $\frac{\omega^2}{\omega^2 + 4n^2\pi^2T^2_c}$ for $n\in\{1,...,7\}$
over $P(d\omega)$ yields explicit upper and lower bounds on the electron-phonon coupling constant $\lambda$, given by
$1/\mathfrak{k}^{(4)}(P,T_c)$ and $1/\mathfrak{k}^*(P,T_c)$, respectively.
\end{quote}

 Our $T_c$ bounds accomplish the following regarding the existing literature, where traditionally $T_c$ has been 
investigated through a linear stability analysis of the normal state, involving some heuristic truncation 
to a finite-dimensional matrix problem of the linearized Eliashberg gap equation in their original formulation.
 First of all, our $T_c$ bounds are obtained rigorously, which in itself is a novelty in this area of research.
 Second, the {closed form} lower bounds for $N\in\{3,4\}$, and our explicit upper bound are new.
 Third, the {closed form} lower bounds for $N\in\{1,2\}$ {vindicate the $T_c$ bounds claimed} in \cite{AllenDynes},
as discussed next.

 Our lower bound (\ref{eq:TcONE}) on $T_c$ obtained with $N=1$, and given as inverse function of the lower bound 
(\ref{eq:kONE}) on $\mathfrak{k}(P,T)$, agrees with the lower bound stated in \cite{AllenDynes} in their formula (19) 
when their $\mu^*(N)=0$.
 It is also mentioned in \cite{AllenMitrovic} and \cite{Carbotte}.
 Of course, this is unsurprising, for a truncation of the linearized Eliashberg gap equation to the first positive Matsubara 
frequency inevitably yields this result, no matter which formulation one uses for the full theory.
 Allen and Dynes start from eq.(7) of \cite{BergmannRainer}.

 Our second-lowest bound $T_c^{(2)}(\lambda,P)$ on $T_c(\lambda,P)$, given as inverse function (w.r.t. $T$) of the lower bound 
(\ref{eq:kTWO}) on $\mathfrak{k}(P,T)$, and which is $>0$ on the domain $(\lambda_2,\infty)$,
 agrees with what one gets by setting the expression for $\rho_1$ {(and $\mu^*(N)$)} in formula (23) 
of \cite{AllenDynes} equal to zero, demanding the vanishing of this largest eigenvalue of an auxiliary $2\times 2$ matrix.
 However, unlike Allen and Dynes we actually proved that the resulting expression is a lower bound to $T_c$, by proving
the invertibility of $T\mapsto \mathfrak{k}^{(2)}(P,T)$, see Proposition~9 in section~\ref{sec:TcB}.

 We also found that it is not correct when \cite{AllenDynes} claim that from their (23) 
``one can generate an explicit lower bound to $T_c$ analogous to ... (21)'' [that] ``is complicated and uninstructive''
 [and therefore not displayed] (the quotes are from \cite{AllenDynes}, fleshed out here in a manner that 
captures the apparent spirit of their statement).
 The explicit lower bound (21) in \cite{AllenDynes} is for the non-dispersive limit of the Eliashberg model, when
their inequality (19) can indeed be inverted to yield an explicit lower bound for $T_c(\lambda,\Omega)$ 
(shorthand for $T_c(\lambda,P)$ when $P(d\omega)=\delta(\omega-\Omega)d\omega$).
 However, as we noted in section~\ref{sec:TcB}, although invertible, even for the simpler non-dispersive model
 the inverse of the map $T\mapsto\Lambda_{\mbox{\tiny{\rm E}}}(\Omega,T)$ 
cannot be computed explicitly in closed algebraic form ``analogous[ly] to (21),'' contrary to what is claimed in \cite{AllenDynes}.

 Our third-lowest and fourth-lowest bounds on $T_c(\lambda,P)$, based on the largest eigenvalue of a pertinent $3\times 3$, 
respectively $4\times 4$ matrix, have not been stated before --- as far as we are able to tell.
 They are given in terms of their inverse functions $T\mapsto\Lambda^{(N)}(P,T)$ in Theorem~4, whose inverses we showed to 
exist for when $T> T_*(P)$; see Proposition~1.
 Once a choice of $P$ has been made, one can carry out the quadratures --- numerically, if necessary --- and 
plot these bounds on $\lambda$ as functions of $T_c$ and visually obtain lower bounds on $T_c(\lambda,P)$.

 Moreover, while Allen and Dynes \cite{AllenDynes} claimed that their increasing sequence of auxiliary eigenvalues yields
an increasing sequence of lower bounds on $T_c$, they did not supply any compelling argument that $T_c$ actually exists,
and why their sequence would converge to it monotonically.
 Their narrative seems to have been largely informed by their numerical studies of the truncated Eliashberg gap equations for the
non-dispersive model, based on which they made sweeping, much more general statements.
 By contrast, we proved that our sequence of lower bounds $T_c^{(N)}(\lambda,P)$ converges upward to $T_c(\lambda,P)$, for each 
$P\in\mathcal{P}$.

Finally, we emphasize that our asymptotically (for $\lambda\sim\infty$) exact upper bound (\ref{eq:CONJbound}) 
on $T_c(\lambda,P)$, that we conjecture to be an upper bound on $T_c(\lambda,P)$ for all $\lambda>0$,
has previously played an important role in the derivation of a physically fundamental upper limit on $T_c$ 
arising from considerations beyond the Eliashberg theory \cite{SemenokBorisEmil,EKS,Tra,Sad}.
 By contrast, we in our series of papers obtain bounds on $T_c$ from within this theory, without considering 
whether or not it by itself provides a valid description of any actual physical system. 
 For such considerations, see 
\cite{YuzAltPRB2,SemenokBorisEmil,EsterlisETal,EKS,CAEK,YuzKieAltPRB,Tra,Sad,YuzAltPat}. 
\newpage

\subsection{Outlook}

 Any more detailed evaluations necessitate the specification of the measure $P(d\omega)$.
 In our next paper, part III of our series, we will specialize the analysis of the present paper to the non-dispersive limit, 
which operates with $P(d\omega) = \delta(\omega-\Omega)d\omega$, with Einstein phonon frequency $\Omega>0$.
 In this case all of the quadratures involving $P$ can be carried out explicitly and much more detailed information can be obtained
by evaluation of our general formulas obtained in the present paper. 
 Having the detailed information for the non-dispersive model one also will be able to compare with empirical results on $T_c$ when
the electron-phonon interactions in superconductors are mediated by optical phonons. 

\bigskip

{\bf Acknowledgement:} We thank the two referees for their constructive criticisms. We are particularly grateful to the
referee who spotted a mistake in our original proof of Proposition~4, now corrected.
 We also thank Steven Kivelson for his comments. 

\bigskip
\vfill

DATA AVAILABILITY STATEMENT: No data have been produced for this paper.

CONFLICT OF INTEREST STATEMENT: The authors declare that they have no conflict of interest.

\end{document}